\documentclass[runningheads]{llncs}
\let\oldmaketitle\maketitle
\renewcommand{\maketitle}{\onecolumn\oldmaketitle}

\usepackage[year=2026,ID=4449]{eccv}

\usepackage{eccvabbrv}

\usepackage{graphicx}
\usepackage{booktabs}

\usepackage[accsupp]{axessibility}  %

\usepackage[pagebackref,breaklinks,colorlinks,citecolor=eccvblue]{hyperref}

\usepackage{orcidlink}

\usepackage{amssymb}%
\usepackage{pifont}%

\usepackage{enumitem}
\usepackage{symbols}

\usepackage{amsmath,amsfonts,bm}

\def\1{\bm{1}}

\def\sp{space}

\def\vc{{\bm{c}}}

\def\vt{{\bm{t}}}
\def\vu{{\bm{u}}}
\def\vv{{\bm{v}}}

\def\vx{{\bm{x}}}
\def\vy{{\bm{y}}}

\def\mI{{\bm{I}}}

\DeclareMathAlphabet{\mathsfit}{\encodingdefault}{\sfdefault}{m}{sl}
\SetMathAlphabet{\mathsfit}{bold}{\encodingdefault}{\sfdefault}{bx}{n}

\def\gD{{\mathcal{D}}}

\def\gV{{\mathcal{V}}}

\def\sR{{\mathbb{R}}}

\newcommand*{\ShowNotes}{} %
\usepackage{soul}
\usepackage{placeins}
\definecolor{darkred}{rgb}{0.7,0.1,0.1}
\definecolor{darkgreen}{rgb}{0.1,0.7,0.1}
\definecolor{cyan}{rgb}{0.7,0.0,0.7}
\definecolor{dblue}{rgb}{0.2,0.2,0.8}
\definecolor{maroon}{rgb}{0.76,.13,.28}
\definecolor{burntorange}{rgb}{0.81,.33,0}
\definecolor{tealblue}{rgb}{0.212,0.459, 0.533}

\definecolor{mypink}{rgb}{0.93359375, 0.62109375, 0.83984375}

\definecolor{pp}{rgb}{0.43921569, 0.18823529, 0.62745098}
\definecolor{rr}{rgb}{0.5254902 , 0.00784314, 0.12941176}
\definecolor{bb}{rgb}{0.09019608, 0.23529412, 0.37647059}
\definecolor{yy}{rgb}{0.49803922, 0.3372549 , 0.0}
\definecolor{gg}{rgb}{0.02352941, 0.3372549 , 0.17647059}

\ifdefined\ShowNotes
  \newcommand{\colornote}[3]{{\color{#1}\bf{#2: #3}\normalfont}}
\else
  \newcommand{\colornote}[3]{}
\fi

\newcommand{\eat}[1]{} %
\definecolor{mygray}{rgb}{0.8627451 , 0.92941176, 0.96862745}

\newcommand{\myparagraph}[1]{\vspace*{2pt}{\bf #1}}

\newcommand{\name}{{\textit{WebAccessVL}}}

\usepackage[export]{adjustbox} %

\usepackage{multirow}
\usepackage{booktabs} %
\usepackage[table]{xcolor}
\usepackage{mathtools}
\usepackage{wrapfig}
\usepackage{verbatim}
\usepackage{listings}
\usepackage{fancyvrb}
\usepackage{siunitx}

\usepackage[numbers,sort&compress]{natbib}

\author{
Amber Yijia Zheng$^\star$ \hspace{.1cm} \and
Jae Joong Lee$^\star$ \hspace{.1cm} \and
Bedrich Benes \hspace{.1cm} \and
Raymond A. Yeh \hspace{.1cm} 
\institute{Department of Computer Science, Purdue University \\
\email{\tt\{\small\tt zheng709, lee2161, bbenes, rayyeh\}@purdue.edu}
}
}

\begin{document}
\title{
WebAccessVL: Violation-Aware VLM\\ for Web Accessibility
}

\maketitle
\begin{figure}[t]
    \centering
    \includegraphics[width=0.95\linewidth]{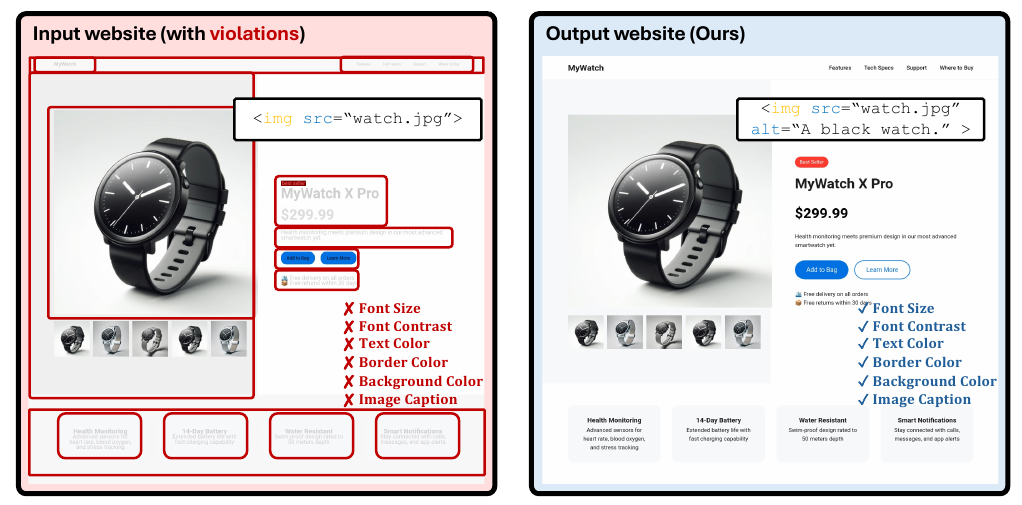}
    \vspace{-0.1cm}
    \caption{Given an input HTML with accessibility violations (visualized by boxes in {\bf \color{BrickRed}red}),~\eg, poor contrast, our method {makes the HTML WCAG2 compliant by refining it to have improved contrast, a better layout, and an appropriate alt-text.} }
    \label{fig:teaser}
    \vspace{-0.2cm}
\end{figure}

\begin{abstract}
    We present a vision-language model (VLM) that automatically edits website HTML to address violations of the Web Content Accessibility Guidelines 2 (WCAG2) while preserving the original design. We formulate this as a supervised image-conditioned program synthesis task, where the model learns to correct HTML given both the code and its visual rendering. We create WebAccessVL, a website dataset with manually corrected accessibility violations. We then propose a violation-conditioned VLM that further takes the detected violations' descriptions from a checker as input. This conditioning enables an iterative checker-in-the-loop refinement strategy at test time. We conduct extensive evaluation on both open API and open-weight models. Empirically, our method achieves 0.211 violations per website, a 96.0\% reduction from the 5.34 violations in raw data and 87\% better than GPT-5. A perceptual study also confirms that our edited websites better maintain the original visual appearance and content. 


\end{abstract}

\section{Introduction}
An accessible web is ``perceivable, operable, understandable, and robust'' to everyone, including people with disabilities
as defined by WCAG2~\cite{w3c_wcag20}. Certain color combinations, such as red and green, may not be distinguished by people with protanopia (partial color blindness) and should be avoided. Web accessibility ensures that everyone has equal access to information and is legally required in the US (through the ADA~\cite{ada_web_guidance}) or in the EU (through the EAA~\cite{eaa_website}). While required by law, surveys by WebAIM~\cite{webaim_million,webaim_screenreader_survey8} found that ``95.9\% of home pages had detected WCAG2 failures'' and that the developers often ``lacked the awareness and skills'' to design WCAG2 compliant websites. We show that advances in VLMs can help tackle these challenges.  

There has been growing interest in leveraging artificial intelligence (AI) to improve web accessibility, for example, by employing large language models (LLMs) to automatically evaluate accessibility compliance~\cite{ahmed2025code}. Other efforts focus on correcting accessibility issues in websites and mobile interfaces by prompting LLMs~\cite{huang2024access,mehralian2024automated}.
These approaches rely exclusively on text-based reasoning \textit{without incorporating any visual information}. That is, they overlook the rendered appearance of a website, the primary way in which users perceive HTML code.
Some accessibility violations are more apparent in the rendered view, and less so in the code. This observation motivates us to integrate website renderings as part of the input to the model.  

We formulate the task of making a website accessible as a computer vision problem of \textbf{image-conditioned program synthesis}. Given an HTML document and a screenshot of its rendering, the goal is to produce a modified HTML of the ``same website'' while ensuring WCAG2 compliance (see~\figref{fig:teaser} for illustration). The main challenges include {\bf (a)} designing a model that is aware of the WCAG2 guidelines and applies appropriate modifications; {\bf (b)} maintaining the output HTML to be faithful to the input website design and rendering; {\bf (c)} developing a suitable benchmark and metrics to quantify models' performance.

In this work, we create a dataset consisting of 1,500 
webpage HTMLs. We then manually correct websites' HTML to better follow WCAG 2.0 standards and use these as ground-truth annotations. Each HTML takes around 7 to 10 minutes to edit. For evaluation, we introduce quantitative metrics to study the accessibility (\# of violations) and faithfulness to the original input (similarity in rendering). Finally, we benchmark existing open-API and open-weight LLMs and VLMs. 

\begin{figure*}[t]
    \centering
    \includegraphics[width=0.99\textwidth]{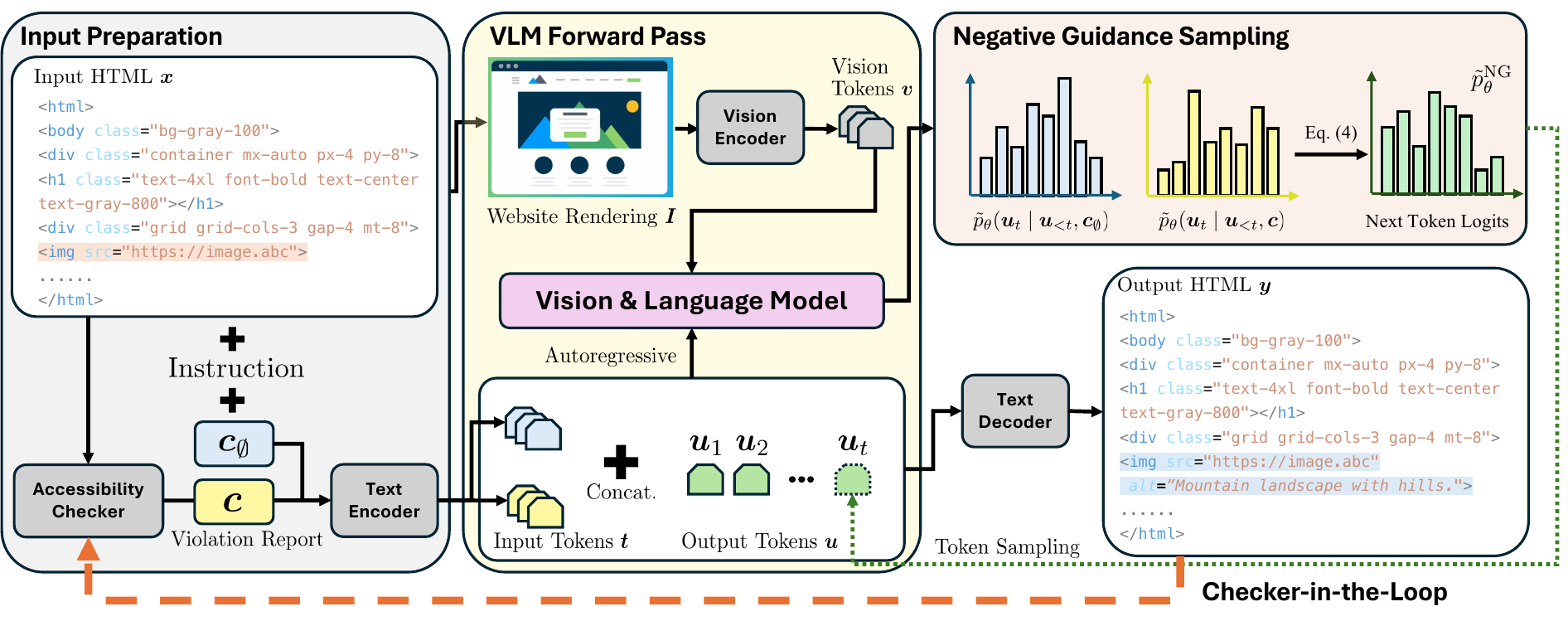}
    \vspace{-0.15cm}
    \caption{\textbf{Overview of our violation-conditioned VLM pipeline.} \textbf{Input Preparation:} The input HTML $\vx$ is analyzed by an accessibility checker to generate a violation report $\vc$, which is encoded with $\vx$ into input tokens $\vt$. \textbf{VLM Forward Pass:} The website rendering $\mI$ is processed by a vision encoder to obtain vision tokens $\vv$. The vision and language model autoregressively generates output tokens $\vu$ conditioned on $\vt$ and~$\vv$. \textbf{Negative Guidance Sampling:} We apply negative guidance sampling to refine the token logits, amplifying the effect of the violation condition $\vc$ to reduce violations in the generated HTML $\vy$. \textbf{Checker-in-the-Loop:} The checker iteratively refines the output by feeding remaining violations back into the model for further correction.
    }
    \vspace{-0.3cm}
    \label{fig:pipeline}
\end{figure*}

For paired data, a baseline is to perform supervised fine-tuning (SFT), \ie, take an HTML document as input and generate a more WCAG2-compliant HTML document. In contrast, we propose training the model to be \textit{violation-aware} by conditioning on parsed violation reports that describe specific issues detected in the input HTML. This conditioning has two key benefits: (a) it guides the model to focus on editing the flagged elements rather than generic accessibility edits that may inadvertently edit unrelated components, and (b) it enables an iterative checker-in-the-loop refinement strategy (at test time) where the model receives feedback on the remaining violations and progressively corrects them. Additionally, we incorporate classifier-free guidance to amplify the effect of violation conditioning; See~\figref{fig:pipeline} for an overview.
We conduct extensive experiments across 17 models, including LLMs and VLMs. Our violation-conditioned VLM achieves 0.211 violations per website (Gemma 3) and 0.244 (Llama 3.2 Vision), an 87\% reduction compared with GPT-5 (1.68 violations), which has the lowest count among baselines. More importantly, we maintain 90\% structural accuracy versus GPT-5's 0.5\%, demonstrating that our method preserves design while fixing accessibility.

{\bf\noindent Our main contributions are as follows}:
\begin{enumerate}[topsep=0pt]
    \item We formulate the task of website accessibility as image-conditioned program synthesis and create a dataset (WebAccessVL) of paired HTMLs for training and evaluation on this task.
    \item We propose a violation-conditioned VLM, which adapts pre-trained VLMs for web accessibility. The design incorporates the violation description into the conditioning and hence enables a test-time refinement approach to further improve performance.
    \item 
    Through extensive experiments, we show that VLMs outperform LLMs for this task and the effectiveness of our approach over baselines.
\end{enumerate}

\section{Related Work}

\myparagraph{AI for accessibility.} 
Traditional accessibility checking relies on rule-based systems~\cite{axe-core,lighthouse}. They require manual maintenance of specialized constraints per violation type, and cannot reason about visual-semantic context, \eg, generating descriptive alt-text from images. To address these challenges, existing work has leveraged AI to tackle accessibility challenges, \eg, automated tools designed to suggest source code fixes %
detected by checkers in mobile apps~\cite{mehralian2024automated}. \citet{ahmed2025code} evaluate ChatGPT's ability to generate and improve web pages in line with WCAG and show that prompting only resolves simple issues, but struggles with complex tasks. Other works have also employed LLMs, such as ChatGPT, to detect~\cite{lopez2024turning} or fix violations through a prompting-based method~\cite{othman2023fostering,aljedaani2024does,huang2024access,andruccioli2025tabular}. %
In contrast, we introduce a violation-conditioned model that is fine-tuned on a dedicated accessibility dataset created in this work. Importantly, we take both visual and textual input. This allows our method to generate HTML that follows the accessibility guidelines while preserving the original design, providing a more effective solution for web accessibility.

\myparagraph{Visual programming and HTML synthesis.} 
Early work by~\citet{andreas2016neural} tackles the problem of visual question answering by assembling neural modules to build an end-to-end differentiable network, one of the first works to output a program from an input image. More recently, VisProg~\cite{gupta2023visual} proposes an approach that tackles complex, compositional visual tasks when given natural language instructions and outputs a program that may call other off-the-shelf computer vision models to perform the task, demonstrating impressive zero-shot reasoning capability. %
In contrast, this work focuses on generating accessible HTML while preserving the website design. 

Research on HTML synthesis primarily focuses on converting rendered website images into code. 
\citet{beltramelli2018pix2code} introduce an end-to-end method for generating code from graphical user interface screenshots using CNNs and RNNs. Building upon VLMs, \citet{lee2023pix2struct} leverage a pre-trained VLM to transform webpage screenshots into simplified HTML. More recently, the synthetic WebSight dataset~\cite{websight} provides HTML code paired with corresponding screenshots to facilitate research in this domain, and~\citet{image2struct} introduced a benchmark for evaluating VLMs on extracting structural information from images, including HTML code. Recent studies present datasets and benchmarks for HTML code synthesis using VLMs~\cite{web2code,interaction2code}, but these works do not focus on accessibility.  

\myparagraph{Vision-Language Models} have recently achieved remarkable progress by jointly learning from visual and textual data. A typical VLM architecture consists of a visual encoder, a vision-language connector, and a large language model (LLM). For instance, LLaVA~\cite{liu2023llava,liu2024llava15} is an open-source multimodal framework that leverages the language-only GPT-4 to generate instruction-following data in a multimodal setting and integrates a CLIP-based vision encoder with the LLM. 
 
Next, Qwen-VL Series~\cite{wang2024qwen2vl,qwen25vl} supports multiple languages and can handle multiple images concurrently during training, building on the Qwen language models~\cite{yang2024qwen2}. Many other deep-net architectures have also been proposed for VLMs. Models such as OtterHD~\cite{li2023otterhd}, mPLUG-Owl~\cite{ye2024mplug}, and InternLM-XComposer2-4KHD~\cite{dong2024internlmxcomposerkhd} are designed to utilize high-resolution images, while LLaVA-NeXT~\cite{liu2024llava16}, Mini-Gemini~\cite{li2024mini}, and MM1~\cite{mckinzie2024mm1} divide the input image into multiple crops. 
We show that prompting off-the-shelf VLMs yields some gains in accessibility compliance. In contrast, SFT on our proposed dataset, \name, combined with the negative guidance, significantly improves the model’s ability to correct accessibility violations.

\section{Approach}
Our goal is to build a model that refines the HTML of a website to be WCAG2-compliant while maintaining the website's content and design. To accomplish this, we propose a violation-conditioned VLM that adapts a pre-trained model for this task (\secref{sec:vc-vlm}). We then describe the dataset collected for SFT (\secref{sec:dataset}).

\subsection{Violation-conditioned VLM}\label{sec:vc-vlm}
{\noindent\bf Problem formulation.} 
Given the input HTML of a website~$\vx$ and its rendering~$\mI$, the model $p_\theta$ aims to output a
revised HTML~$\vy$ of the same website but with fewer accessibility violations under WCAG2. To guide the repair process, we condition the model on a parsed violation report $\vc$ that describes the specific WCAG2 violations detected in the input HTML $\vx$. This report is obtained by running an accessibility checker (\eg, IBM's checker~\cite{checker}) on $\vx$, which identifies violation types, affected elements, and their locations in the code. Conceptually, the model is an image-conditioned program generator. The model is trained by minimizing the negative log-likelihood over the training set $\gD$, \ie,
\bea
\min_\theta \sum_{(\vx, \mI, \vy, \vc)\in \gD} -\log p_\theta(\vy|\vx, \mI, \vc).
\eea
Next, we describe how to adapt $p_\theta$ from a pre-trained VLM. An overview of the approach is shown in~\figref{fig:pipeline}.

\myparagraph{VLM details.} Our violation-conditioned VLM is adapted from a pre-trained autoregressive token-based model, in which textual and visual inputs are tokenized by text/vision encoders. The model then autoregressively generates a sequence of output tokens, which are subsequently decoded into text.

To prepare the inputs for the VLM, we construct a prompt that includes: (i) a structured violation report $\vc$ listing the violations detected in the input HTML $\vx$, (ii) the instruction ``\texttt{The expected output HTML has zero violations}'', and (iii) the input HTML $\vx$ itself. The entire prompt is encoded into text tokens $\vt = (\vt_1, \vt_2, \dots, \vt_L)$. The violation report $\vc$ is serialized into a structured text format describing each violation with its rule ID, message, and location (\eg, ``\texttt{Violation 1: text\_contrast\_sufficient, Issue: Text contrast is 2.1:1, Location: //div[@id='header']}'').  Without the prepended instruction, we found that the model's outputs are likely not in HTML format. Next, the rendered image~$\mI$ is processed by the vision encoder to obtain vision tokens~$\vv = (\vv_1, \dots, \vv_K)$. 
During training, $\vc$ describes the violations detected in the input HTML $\vx$, and the model learns to generate the corrected HTML $\vy$ that fixes these violations.
The target HTML $\vy$ is also encoded into text tokens $\vu = (\vu_1, \vu_2, \dots, \vu_T)$. SFT of the 
VLM $\tilde{p}_\theta$ is performed by minimizing the token-level negative log-likelihood:
\bea
\min_\theta \sum_{(\vt, \vv, \vu, \vc)\in \tilde{\gD}} \sum_{t=1}^T -\log \tilde{p}_\theta(\vu_t \mid \vu_{<t}, \vt, \vv, \vc),
\eea
where $\tilde{\gD}$ denotes the tokenized training set.

\myparagraph{Negative guidance sampling.} 
To generate an HTML without violations~$\vy$, we condition on the detected violation report $\vc$ from the input HTML, sample from~$\tilde{p}_{\theta}$, and decode the set of generated tokens. In practice, we found that using a more advanced sampling strategy further reduces the number of violations. Specifically, we incorporate LLM classifier-free guidance~\cite{rombach2022high,du2020compositional,ho2022classifier,sanchez2023stay}. Let
\bea
\ell_\theta(\vu_t \mid \vu_{<t}, \vc) \triangleq \log \tilde{p}_\theta(\vu_t \mid \vu_{<t}, \vt, \vv, \vc)
\eea
denote the conditional logit for the token $\vu_t$. 
During training, we randomly drop the violation report with probability $p_{\text{uncond}}=0.1$ to enable classifier-free guidance. At inference, we use negative guidance sampling to amplify the effect of conditioning on violations. 

Formally, the negative guided logit for the token $\vu_t$ is
\bea
\label{eq:cfg}
\ell^{\text{NG}}_\theta(\vu_t, \gamma) \triangleq \ell_\theta(\vu_t \mid \vu_{<t}, \vc_{\emptyset}) + \gamma \cdot \Bigl( \ell_\theta(\vu_t \mid \vu_{<t}, \vc) - \ell_\theta(\vu_t \mid \vu_{<t}, \vc_{\emptyset})\Bigr),
\eea
where $\gamma \in \sR_+$ is the guidance scale, $\vc$ denotes the violation report from the input HTML, and $\vc_{\emptyset}$ denotes an unconditional prompt that instructs the model to return the HTML as-is without modifications. Intuitively, when $\gamma=1$, the sampling reduces to sampling from the conditional distribution with $\vc$. For $\gamma > 1$, the model amplifies the effect of conditioning on the specific violations, upweighting tokens that lead to corrected HTML and downweighting tokens associated with unchanged output. We explored alternative unconditional prompt formulations and empirically determined this configuration to be most effective (see~\secref{sec:ablation_study}).

The final probability of token $\vu_t$ is then computed via a Softmax over the vocabulary set $\gV$, defined as:
\bea
\tilde{p}^{\text{NG}}_\theta(\vu_t \mid \vu_{<t}, \vt, \vv, \vc) =
 {\exp\Bigl(\ell^{\text{NG}}_\theta(\vu_t, \gamma)\Bigr)}\Big/{\sum_{\vu' \in \gV} \exp\Bigl(\ell^{\text{NG}}_\theta(\vu', \gamma)\Bigr)}.
\eea
Finally, the sampled output tokens $\vu$ are fed to the text decoder to generate the revised HTML $\vy$. Note, extra text is occasionally generated outside the HTML structure. Hence, we apply postprocessing to extract the HTML segment using a non-greedy regular expression that matches the content enclosed between the \texttt{<html>} and \texttt{</html>} tags.

\myparagraph{Checker-in-the-loop inference.}
The violation conditioning mechanism described above enables an iterative refinement strategy at inference time. Let $\mathcal{C}(\cdot)$ denote the accessibility checker that maps HTML to its corresponding violation report. We define an iterative process where, at iteration $i$, we generate:
\bea
\vy^{(i)} \sim p_\theta(\cdot \mid \vx^{(i-1)}, \mI^{(i-1)}, \vc^{(i-1)}), \quad \vc^{(i)} = \mathcal{C}(\vy^{(i)}),
\eea
with initial conditions $\vx^{(0)} = \vx$, $\mI^{(0)} = \mI$, and $\vc^{(0)} = \mathcal{C}(\vx)$. This process continues for $i = 1, \dots, K$ or until convergence, defined as either $|\vc^{(i)}| = 0$ or $\vy^{(i)} = \vy^{(i-1)}$, where $|\vc^{(i)}|$ denotes the number of violations. %

Crucially, without violation conditioning, the checker's output cannot be meaningfully fed back into the model; in this case, the model would simply receive the same HTML input repeatedly without guidance on what remains to be fixed. By conditioning on the specific violations detected at each iteration, the model receives targeted feedback on which accessibility issues persist and can better focus its corrections accordingly. %
In practice, we find that most samples converge in the second pass, with minimal improvement from additional passes.

\subsection{WebAccessVL Dataset}\label{sec:dataset}
To our knowledge, no public datasets are available for training models to refine webpage HTML to follow accessibility guidelines. To address this issue, we collect a dataset and will make it publicly available. 

\myparagraph{Dataset construction methodology.}
To construct a set of websites' HTML, we randomly sampled 1,500 websites from a large-scale HTML dataset provided by~\citet{websight}. During sampling, we verify the availability of essential assets (\eg, images and icons). Websites with missing assets that led to rendering errors were excluded. To ensure reproducibility, we have saved all required assets locally.

To annotate the HTML files, an annotator modifies the code and repeatedly renders each webpage to ensure that its visual design remains consistent with the original version. We use an industrial-grade accessibility checker developed by IBM~\cite{checker} to minimize accessibility violations to the best of the annotator's ability. {\it On average, this process takes approximately 7 to 10 minutes per webpage.} Notably, due to the technical complexity of the task, the annotator requires an advanced degree in computer science.

\myparagraph{Dataset statistics.}
We analyzed the distribution of the collected websites, focusing on the types of accessibility violations and their dependence on visual versus textual information. Among the 26 violation categories, we identified 8 that correspond to vision-related tasks requiring visual understanding. Our analysis shows that \textit{35.8\% of violations involve vision-related factors}, while the remaining 64.2\% are purely language-based, indicating that visual information should not be overlooked. Below, we briefly describe the major violation types; additional details are provided in Appendix~\secref{supp_sec:violation_details}.

\begin{figure}[t]
\centering
\begin{tabular}{ccc}
\includegraphics[width=0.325\linewidth]{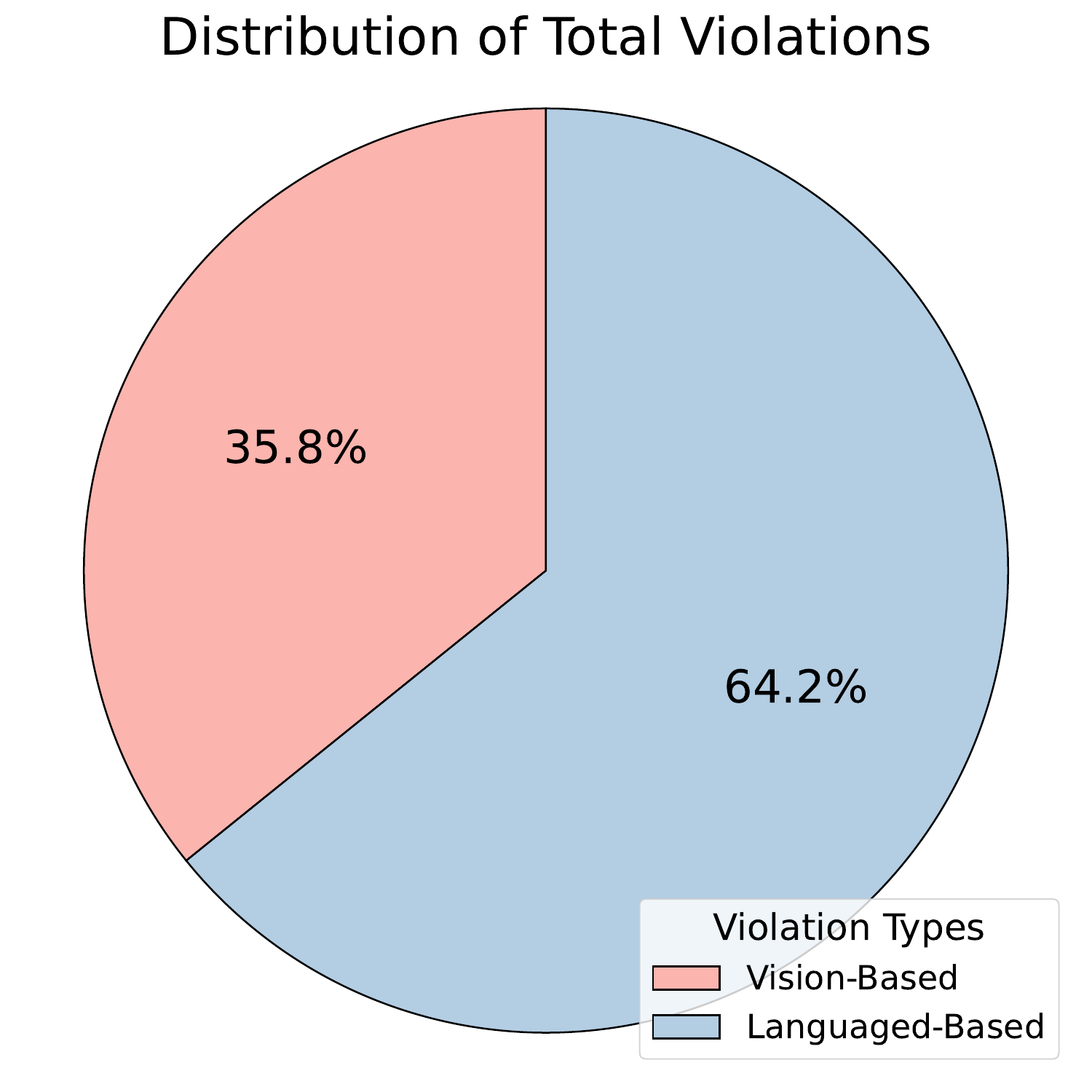} &
\includegraphics[width=0.325\linewidth]{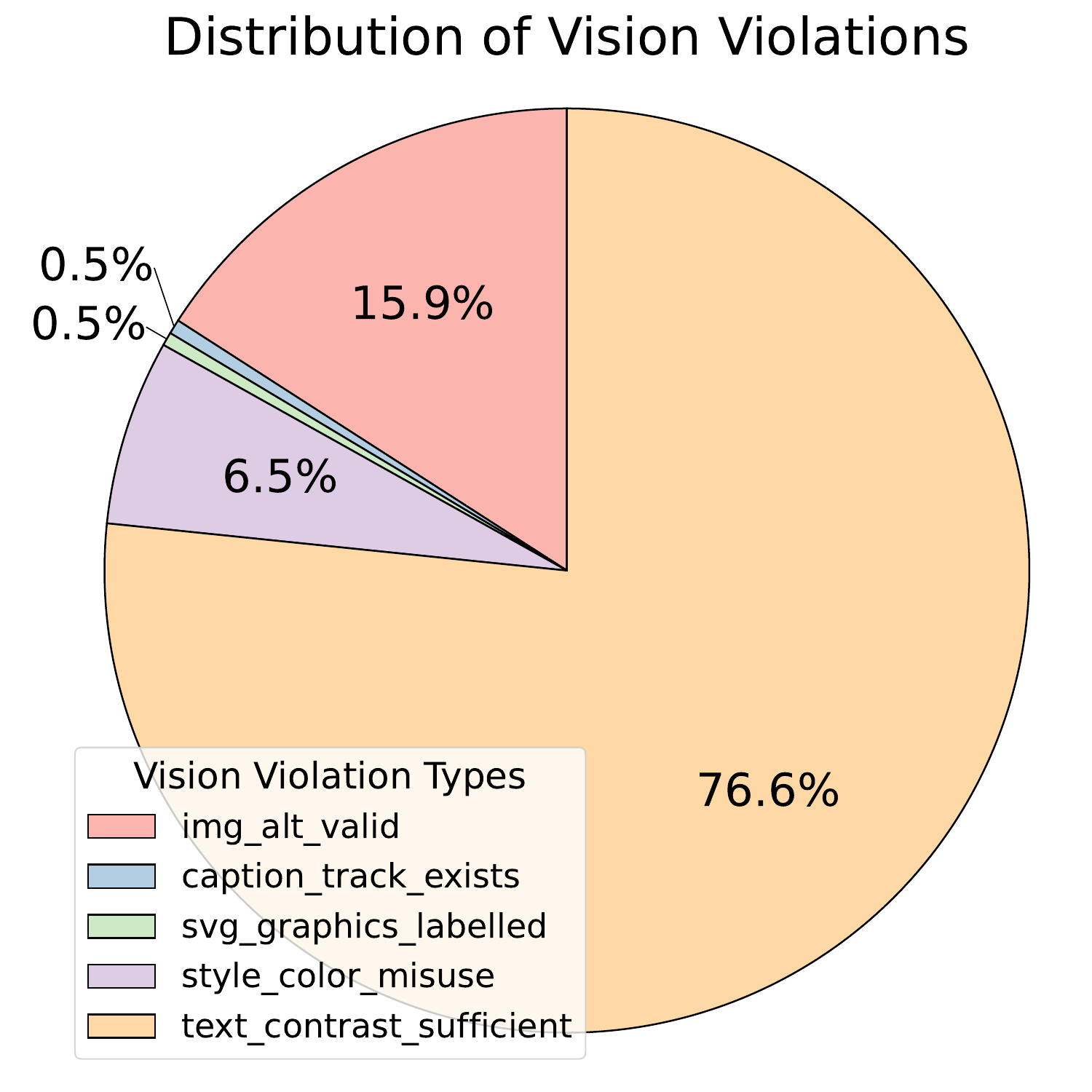}&
\includegraphics[width=0.325\linewidth]{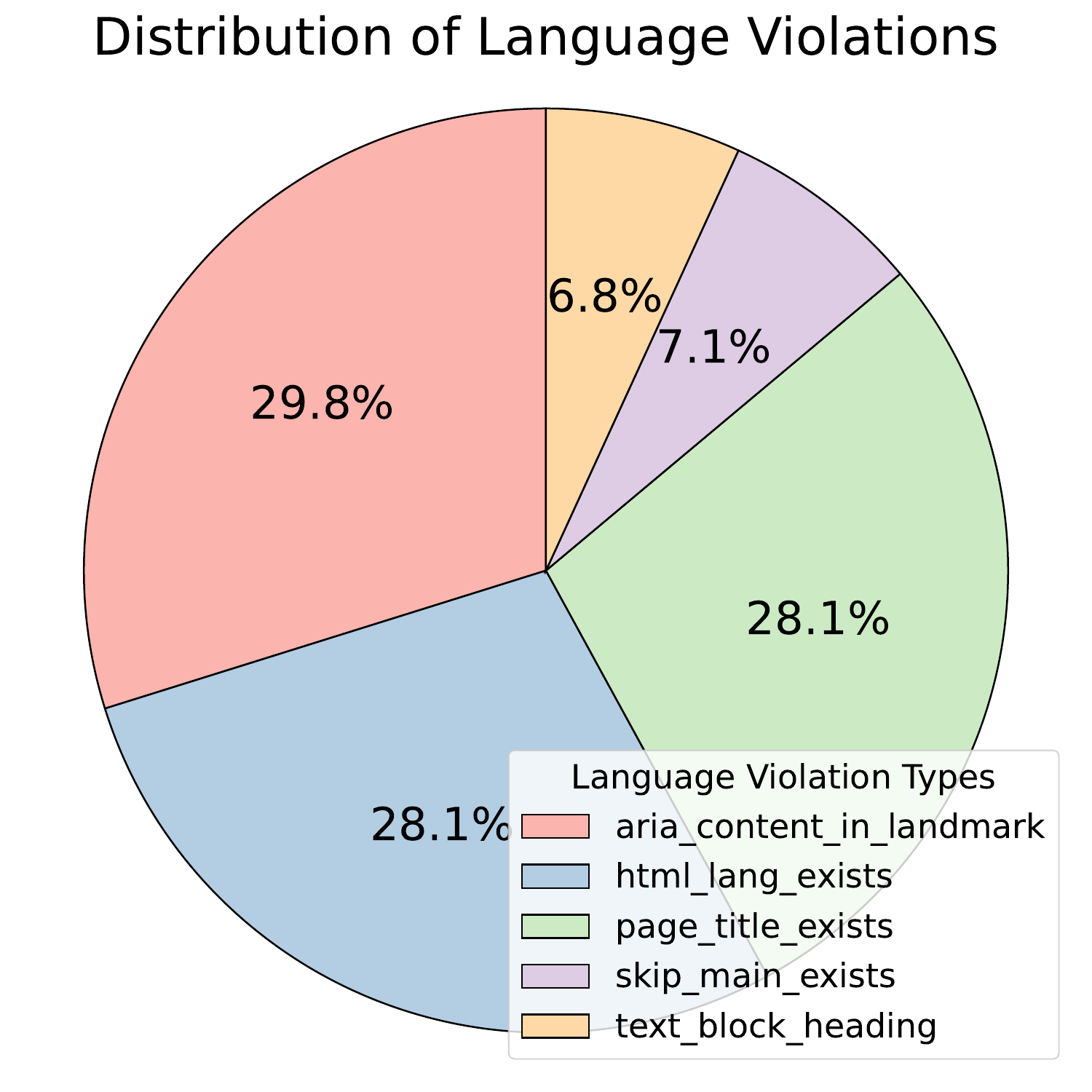} 
\end{tabular}
\vspace{-0.3cm}
\caption{Pie charts showing the distribution of total violations, vision violations, and language violations. While the majority of the violations are language-based, more than a third of the total violations are vision-based. 
The most common vision violation is insufficient text contrast (``text contrast sufficient''), and the most common language violation is ``aria content in landmark''.
}
\label{fig:frequent_violations} 
\vspace{-0.2cm}
\end{figure}

\myparagraph{Vision violations.} 
\figref{fig:frequent_violations} shows the five most frequent vision-related accessibility issues in the dataset, ranked by occurrence rate: {\ding{172}} \texttt{Text contrast sufficient} violation means that the text-to-background contrast ratio is less than 3:1. {\ding{173}} \texttt{Img alt valid} violation is triggered when alternative text is missing.~\ding{174} The checker flags a \texttt{style color misuse} violation when there are only color differences to mark required fields, \ie, users who cannot perceive the color will not be able to perform further actions.~\ding{175} \texttt{caption track exists} violation occurs when there is no textual information for video content.~\ding{176} \texttt{SVG graphics labeled} violation is similar to a missing alt text, but it differs in terms of format.

\myparagraph{Language violations.} 
As shown in~\figref{fig:frequent_violations}, the top five occurring language-related violations in the dataset are: {\ding{182}} \texttt{aria content in landmark} violation is when there is no role or subsection specified.~\ding{183} \texttt{html lang exists} violation is triggered when no language is specified on the website.~\ding{184} \texttt{page title exists} violation is marked if there is no \texttt{head} element and \texttt{title} element in HTML code.~\ding{185} \texttt{skip main exists} violation happens when \texttt{main} tag is missing.~\ding{186} \texttt{text block heading} violation is when there are no headings, \ie, the website has dense blocks of text.

\section{Experiments}

We evaluate our violation-conditioned VLM on the WebAccessVL dataset.
We benchmark open API and open weight models, then evaluate supervised fine-tuning on our dataset. 
Ablation studies validate our design choices, and a perceptual study checks whether quantitative metrics match human judgment.

\subsection{Experimental Setup}

\myparagraph{Dataset.}
We evaluate all models on the test set of WebAccessVL, which contains 1,000 manually curated HTML screenshot pairs.
The training set consists of 500 paired examples covering diverse website types, including e-commerce, news, education, and social media platforms. 
Each HTML is rendered at 1,920$\times$1,080 resolution using a headless Chrome browser. 

\myparagraph{Evaluation.}
We measure whether models improve accessibility while preserving the original design. We report four metrics:
\begin{enumerate}[topsep=0pt]
    \item \textit{Accessibility guidance violations} $(\downarrow)$:
    Average number of WCAG2 violations detected by IBM's accessibility checker~\cite{checker}.
    \item \textit{Caption-Image CLIP score} $(\uparrow)$: We compute the CLIP score between the generated \texttt{Alt} text and the corresponding image to assess the semantic alignment of text and visual content. 
    \item \textit{Structural accuracy} $(\uparrow)$: We render both the output HTML and the ground-truth, then compute the Structural Similarity Index Measure~\cite{wang2004image} (SSIM) to determine if any significant design modifications have occurred. We say a website is structurally consistent if the SSIM value exceeds 0.9. 
    \item \textit{Tree edit distance} $(\downarrow)$: Tree edit distance between input and output HTML. 
    Using the DOM structure, we treat each HTML tag (\texttt{<div>, <body>, <nav>}) as a tree node. Insert, update, and remove operations each cost one. 
\end{enumerate}

\subsection{Main Results}

We compare our approach against 17 baseline models across three categories: (1)~commercial API models (GPT-4o~\cite{GPT4o}, GPT-5~\cite{GPT5}, Claude 3.5 Sonnet~\cite{claude35sonnet}, Gemini 1.5 Flash~\cite{team2024gemini}), (2) open source LLMs (Llama 3.2~\cite{dubey2024llama}, Qwen 2/2.5/3~\cite{yang2024qwen2,qwen2.5,qwen3}), and (3) open source VLMs (Llama 3.2 Vision~\cite{dubey2024llama}, Qwen 2/2.5/3 VL~\cite{wang2024qwen2vl,qwen25vl,qwen3vl}, LLaVA 1.5~\cite{liu2024llava15}/1.6~\cite{liu2024llava16}, DeepSeek VL 2~\cite{wu2024deepseekvl2}, Janus Pro~\cite{chen2025janus}).
All baselines are prompted with detailed instructions on WCAG2 guidelines and asked to modify the HTML to improve accessibility while maintaining visual consistency.

\tabref{tab:main_table} presents the main results. The raw input websites average 5.335 violations, consistent with WebAIM~\cite{webaim_million} findings on widespread accessibility issues.

\myparagraph{The accessibility preservation tradeoff.}
API models show a tradeoff between reducing violations and preserving design. GPT-5 reduced violations to 1.681 but rebuilds websites from scratch. Structural accuracy is near zero (0.005), and tree edit distance is high (15.573). GPT-5 appears to interpret the task as ``generate an accessible website'' rather than ``fix this specific website.'' GPT-4o Vision and Claude 3.5 maintain better structure (0.408 and 0.183 accuracy) but have more violations (2.918 and 2.310). Gemini 1.5 struggles on both.

\myparagraph{Open-source VLMs fall behind APIs.}
Pre-trained open-source VLMs (without fine-tuning) perform poorly, retaining nearly all input violations (4 to 5 violations). Three reasons: (1) these models lack exposure to HTML rendering paired data during pre-training, (2) accessibility guidelines are specialized domain knowledge rarely seen in web corpora, and (3) the models struggle with long HTML documents. Even large recent VLMs like Qwen 3 VL (5.227 violations) barely improve over raw input.
Text-only LLMs show comparable or worse performance than their VLM counterpart when used in the zero-shot manner (Qwen 3: 5.185 vs 5.227 violations). In the SFT setting, VLMs outperform the text-only LLMs; see~\tabref{tab:llm_vs_vlm}.%

\myparagraph{Our method.}
Our fine-tuned models reduce violations by 10$\times$. Llama 3.2 Vision reaches 0.244 violations (95.4\% reduction from raw), 85.5\% better than GPT-5 despite being smaller. However, structural accuracy is low (0.018), comparable to pre-trained models, indicating that the SSIM threshold of 0.9 may be too strict for accessibility modifications. The higher tree edit distance (11.642 vs GPT-4o's 4.939) reflects that accessibility fixes require HTML modifications (\eg, adding ARIA labels, restructuring for landmarks), but these changes preserve visual appearance as shown in our perceptual study. The CLIP scores (0.241) confirm that the generated alt text accurately describes images.
\begin{table*}[t]
    \small
    \centering

    \caption{We show the evaluation metrics using the best open models and our violation-conditioned VLMs on the test set of \name. We present open APIs, open-source LLMs, and open-source VLMs.
    }
            \vspace{-0.2cm}
            \resizebox{\linewidth}{!}
    {%
    \begin{tabular}{ccccccc}
        \specialrule{.15em}{.05em}{.05em}
        & \textbf{Base Model}  &  VLM & \# Violation $\downarrow$   &  Caption-Img CLIP $\uparrow$ & Struct. Acc. $\uparrow$ &Tree Edit Dist. $\downarrow$\\
        \midrule
        & Raw Data & - & 5.335  & - &  - & -\\ \hline
        \multirow{4}{*}{\rotatebox[origin=c]{90}{Open API}} &GPT-4o & \ding{53}  & 2.929   & 0.227 & 0.389 & 5.081\\
        &GPT-5 & \ding{53}  & 1.681   & 0.144 & 0.005 & 15.573\\
        &Claude3.5 Sonnet& \ding{53}& 2.310   & 0.223  & 0.183 & 7.603\\
        &Gemini 1.5 Flash& \ding{53}& 6.213   & 0.229 & 0.286 & 5.762\\
        \cline{1-7}
        
        \multirow{12}{*}{\rotatebox[origin=c]{90}{Open-weight Models}} &Llama 3.2 3B& \ding{53}& 4.382    & 0.173 & 0.020 & 14.097\\
        &Qwen 2 7B & \ding{53}& 5.173  & 0.178 & 0.022 & 1.235\\ 
        &Qwen 2.5 7B & \ding{53}& 5.303    & 0.134 & 0.019 & 0.314\\
        &Qwen 3 8B & \ding{51} & 5.185 & 0.216 & 0.021 & 0.567 \\

        \cline{2-7}
        &GPT-4o Vision & \ding{51} & 2.918    & \bf 0.250 & \bf 0.408 & 4.939 \\
        &DeepSeek VL 2 small& \ding{51} & 5.505   & 0.149 & 0.022 & 2.090\\
        &Janus Pro 7B& \ding{51} & 5.335   &  0.125 & 0.022 & 12.616\\
        &Llama 3.2 Vision 11B& \ding{51} & 4.765   & 0.165 & 0.019 & 15.474\\
        &LLaVA 1.5& \ding{51} & 4.944   & 0.156 & 0.021 & 4.443\\
        &LLaVA 1.6& \ding{51} & 5.203  & 0.169 & 0.022 & 1.095\\
        &Qwen 2 VL 7B& \ding{51} & 5.214   & 0.153 & 0.022 & \bf 0.125\\
        &Qwen 2.5 VL 7B & \ding{51}& 3.770   & 0.217 & 0.019 & 3.983\\ 
        &Qwen 3 VL 8B& \ding{51}& 5.227 & 0.235 & 0.022 & 0.412 \\
        &Gemma 3 12B & \ding{51}& 4.427 & 0.233 & 0.021 & 0.313\\
        \hline

        \cline{2-7}
        \rowcolor{mygray} 
        & LLaVA 1.6 & \ding{51}& 0.366 & 0.236 & 0.019 & 11.141\\
        \rowcolor{mygray} 
        &  Llama 3.2 Vison& \ding{51}& 0.244 & 0.241 & 0.018 & 11.642\\
        \rowcolor{mygray} 
        \multirow{-3}{*}{\rotatebox[origin=c]{90}{Ours}}
        & Gemma 3 12B & \ding{51} & \bf 0.211 & 0.239 & 0.019 & 11.761 \\
        \specialrule{.15em}{.05em}{.05em}
    \end{tabular}
    }

    \label{tab:main_table}
    \vspace{-.5cm}
\end{table*}

\myparagraph{Impact on disability groups.}
We further analyze how our method impacts specific disability groups following the categorization by~\citet{droutsas2025web} shown in~\tabref{tab:disability-impact}. Overall, our method fixes 98.2\% of violations affecting%
\begin{wraptable}[8]{r}{0.45\textwidth}
\centering
\vspace{-1cm}
\caption{Violation fixes by disability group~\cite{droutsas2025web} on 1,000 test samples.}
\label{tab:disability-impact}
\scriptsize
\begin{tabular}{@{}lrrr@{}}
\toprule
\textbf{Group} & \textbf{Before} & \textbf{After} & \textbf{Fix Rate} \\
\midrule
Blind/Low-vision & 2,467 & 45 & 98.2\% \\
Motor/Keyboard & 1,334 & 24 & 98.2\% \\
Low vision/Dyslexia & 1,525 & 142 & 90.7\% \\
\midrule
\textbf{Overall} & \textbf{5,326} & \textbf{211} & \textbf{96.0\%} \\
\bottomrule
\end{tabular}
\end{wraptable}
blind/low-vision users (2,467 violations: missing alt text, ARIA labels, page titles) and motor-impaired users (1,334 violations: keyboard navigation, skip links, landmark structure). Text contrast violations affecting dyslexia and low-vision users have a 90.7\% fix rate (1,383/1,525). This is lower because some fixes require changing the original design. The remaining 4\% involve rare custom ARIA roles that appear in <1\% of training data.

\subsection{VLMs vs. LLMs Comparison}
We compare LLMs and VLMs using the same base architecture to evaluate the importance of visual information.
\tabref{tab:llm_vs_vlm} shows results for GPT-4o, Llama 3.2, and Qwen 3 in text-only (LLM) and vision-enabled (VLM) configurations, with and without SFT.

\begin{table*}[t]
    \small
    \centering

    \caption{Comparison between text-only LLMs and VLMs with and without supervised fine-tuning on WebAccessVL.
    }
            \vspace{-0.2cm}
            \resizebox{\linewidth}{!}
    {%
    \begin{tabular}{ccccccc}
        \specialrule{.15em}{.05em}{.05em}
        \textbf{Base Model}  &  VLM & SFT & \# Violation $\downarrow$   &  Caption-Img CLIP $\uparrow$ & Struct. Acc. $\uparrow$ &Tree Edit Dist. $\downarrow$\\
        \midrule
       Raw Data & - & - & 5.335  & - &  - & -\\ \hline
       \multirow{2}{*}{GPT 4o} & \ding{53} & \ding{53} &  2.929   & 0.227 & 0.389 & 5.081\\
        & \ding{51} & \ding{53} & 2.918    & 0.250 & \bf 0.408 & 4.939\\ \hline
        \multirow{4}{*}{Llama 3.2} & \ding{53} &\ding{53} &  4.382    & 0.173 & 0.020 & 14.097\\
        & \ding{51} & \ding{53} &  4.765   & 0.165 & 0.019 & 15.474 \\
        & \ding{53} & \ding{51} &  1.119   & 0.208 & 0.255 &  3.104 \\
        & \ding{51} & \ding{51} & \bf 0.451   & \bf 0.260 & 0.331 & 3.004 \\ \hline
        \multirow{4}{*}{Qwen 3} & \ding{53} & \ding{53} & 5.185 & 0.216 & 0.021 & 0.567\\
        & \ding{51} & \ding{53} & 5.227 & 0.235 & 0.022 & \bf 0.412\\
        & \ding{53} & \ding{51} & 2.670 & 0.176 & 0.021 & 6.661 \\
        & \ding{51} & \ding{51} &  0.604 & 0.239 & 0.019 & 11.807 \\
        \specialrule{.15em}{.05em}{.05em}
    \end{tabular}
    }

    \label{tab:llm_vs_vlm}
    \vspace{-1cm}
\end{table*}

\myparagraph{VLMs and LLMs are comparable without fine-tuning.}
For GPT-4o, adding vision capabilities gives marginal improvements in the zero-shot setting (2.929 $\rightarrow$ 2.918 violations), with slight gains in structural metrics. Generic vision-language pre-training does not equip models to visually diagnose accessibility issues. While the model can ``see'' the page, it does not know which visual patterns indicate violations.

\myparagraph{SFT improves the importance of visual input.}
For Llama 3.2, SFT reduces violations from 4.382 to 1.119 for text only, but adding vision reduces this to 0.451, a 60\% additional improvement. The VLM version achieves better structural preservation (0.331 vs 0.255), showing that visual understanding helps make more targeted modifications. That is, fine-tuning teaches the model to extract accessibility relevant visual features. %

\myparagraph{Visual input helps vision violations.}
Qwen3 shows a clear pattern; without SFT (zero-shot) performance is nearly identical between LLM and VLM (5.185 vs. 5.227). With SFT, the Qwen3 achieves 77\% fewer violations (2.670 vs. 0.604). VLMs excel at vision-dependent issues (contrast, alt text quality) while performing comparably on structural issues (ARIA landmarks, semantic HTML). Lastly, both LLM and VLM show similarly low structural accuracy post SFT (0.021 and 0.019, respectively), indicating that the accessibility modifications required exceed the strict SSIM threshold regardless of whether visual understanding is used.

\subsection{Ablation on Violation-Awareness}
We conduct ablations to evaluate the contribution of our violation conditioning approach.
\tabref{tab:pretrain_vs_sft} compares four configurations: (a) pre-trained model with zero-shot prompting, (b) standard supervised fine-tuning (SFT), (c) SFT with violation conditioning (VC), and (d) VC with iterative checker in the loop refinement.

\begin{table*}[t]
    \small
    \centering

    \caption{Comparison of pre-trained versus SFT on WebAccessVL, with and without violation-conditioned prompting (VC) and checker-in-the-loop refinement (Loop).
    }
            \vspace{-0.2cm}
            \resizebox{\linewidth}{!}
    {%
    \begin{tabular}{cccccc}
        \specialrule{.15em}{.05em}{.05em}
        \textbf{Base Model}  &  Method & \# Violation $\downarrow$   &  Caption-Img CLIP $\uparrow$ & Struct. Acc. $\uparrow$ &Tree Edit Dist. $\downarrow$\\
        \midrule
       Raw Data & - & 5.335  & - &  - & -\\ \hline
        \multirow{4}{*}{Llama 3.2} & Pre-train & 4.765   & 0.165 & 0.019 & 15.474\\
        & SFT & 0.451   & 0.260 & 0.331 & 3.004 \\
        & + VC & 0.337 & 0.241 & 0.018 & 11.687 \\ 
        \rowcolor{mygray} & + VC \& Loop & 0.244 & 0.241 & 0.018 & 11.642 \\ \hline
        \multirow{4}{*}{LLaVA 1.6} & Pre-train & 5.203  & 0.169 & 0.022 & 1.095 \\
        & SFT & 0.706 & 0.235 &  0.020 & 11.780 \\
        & + VC & 0.517 & 0.230 & 0.019 & 11.147 \\ 
        \rowcolor{mygray} & + VC \& Loop & 0.366 & 0.236 & 0.019 & 11.141 \\ \hline
        \multirow{4}{*}{Gemma 3} & Pre-train & 4.427 & 0.233 & 0.021 & 0.313\\
        & SFT & 0.440 & 0.238 & 0.018 & 12.434\\
        & + VC & 0.352 & 0.239 & 0.019 & 11.747 \\ 
        \rowcolor{mygray} & + VC \& Loop & 0.211 & 0.239 & 0.019 & 11.761 \\ 
        \specialrule{.15em}{.05em}{.05em}
    \end{tabular}
    }
    \label{tab:pretrain_vs_sft}
    \vspace{-.6cm}
\end{table*}

\myparagraph{Violation conditioning is effective.}
Adding VC reduces violations by 25--30\% across models (Llama: 0.451 $\rightarrow$ 0.337, LLaVA: 0.706 $\rightarrow$ 0.517). For Llama 3.2, structural accuracy decreases from 0.331 to 0.018, while for LLaVA 1.6 it drops slightly from 0.020 to 0.019. This suggests that violation conditioning leads to more comprehensive accessibility fixes that exceed the strict SSIM threshold of 0.9, though the qualitative results and perceptual study confirm these changes preserve visual appearance. With conditioning, models make targeted edits focused on elements flagged in the violation report.

\myparagraph{Checker-in-the-loop fixes induced violations.}
Iterative refinement gives a small but consistent gain (Llama: 0.337 $\rightarrow$ 0.244, LLaVA: 0.517 $\rightarrow$ 0.366, Gemma: 0.352 $\rightarrow$ 0.211). Structural accuracy remains stable, showing that the iterative process focuses on fixing violations without introducing additional structural changes.
Note, single-pass models may introduce new violations while fixing existing ones. For example, when attempting to fix contrast issues by modifying the background and text colors, the model may simultaneously update both the background and text to be dark, \ie, creating a new low-contrast violation. The checker-in-the-loop catches these violations (induced by the first pass) in the second round. Most samples converge in 2 to 3 iterations. Finally, we observe consistent improvement across several models, demonstrating the generality of our violation conditioning approach.

\begin{figure*}[t]
    \centering
    \setlength{\tabcolsep}{4pt}
    \renewcommand{\arraystretch}{1.0}
    \small
    \begin{tabular}{ccc>{\columncolor{mygray}}c} %
        Input & Claude3.5 & GPT-5 & \bf Ours\\
        \frame{\includegraphics[width=0.233\linewidth]{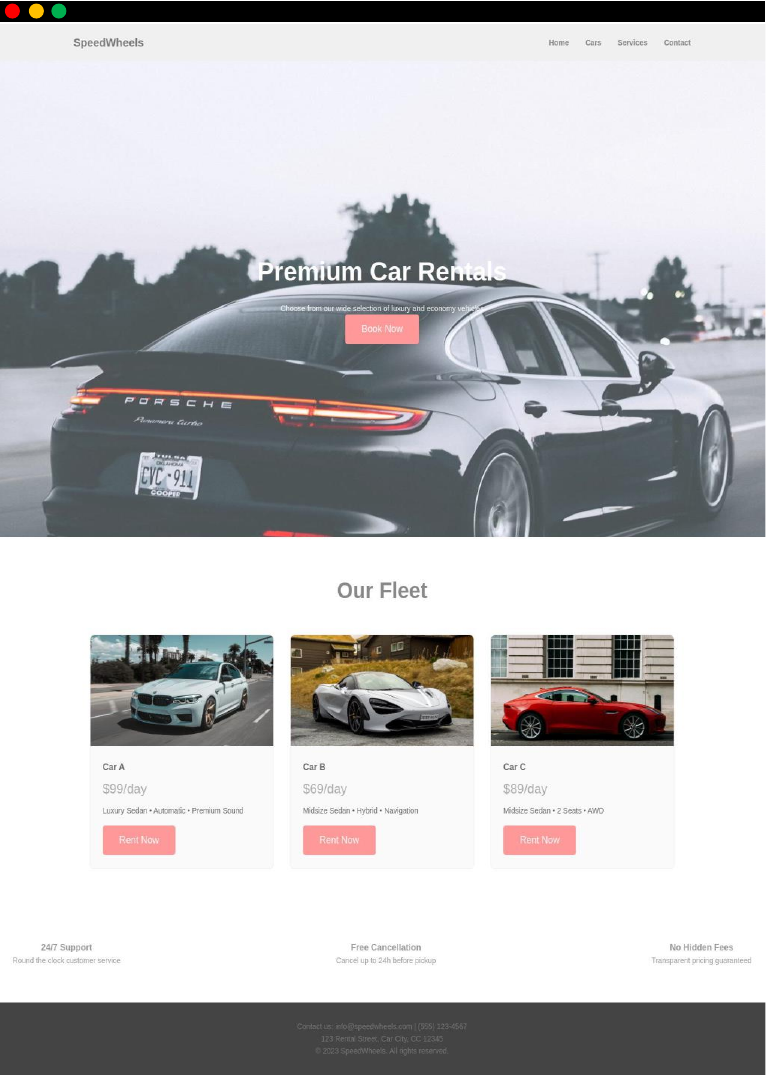}} &
        \frame{\includegraphics[width=0.233\linewidth]{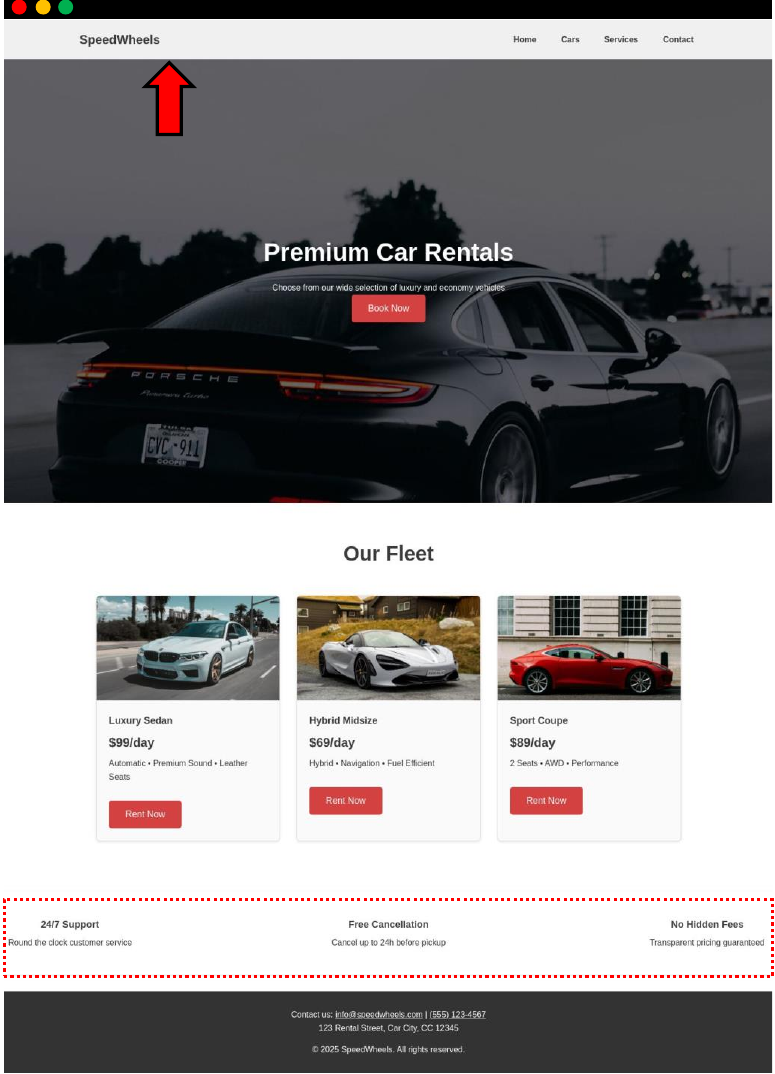}} &
        \frame{\includegraphics[width=0.233\linewidth]{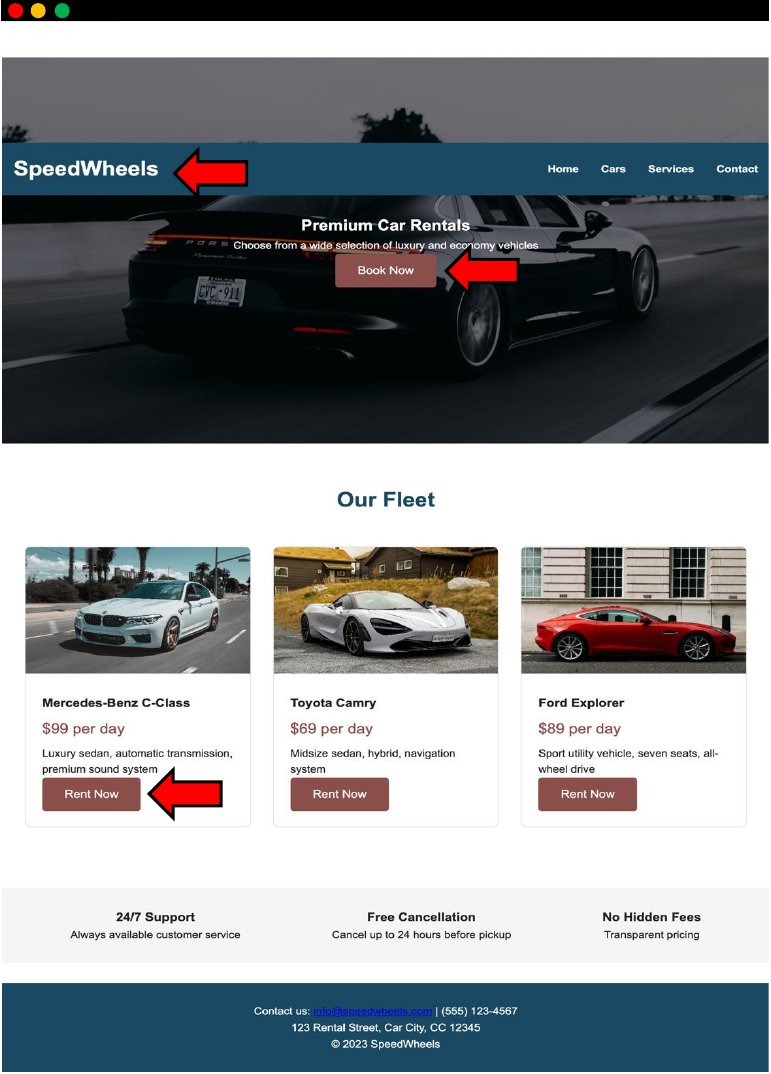}} &
        \frame{\includegraphics[width=0.233\linewidth]{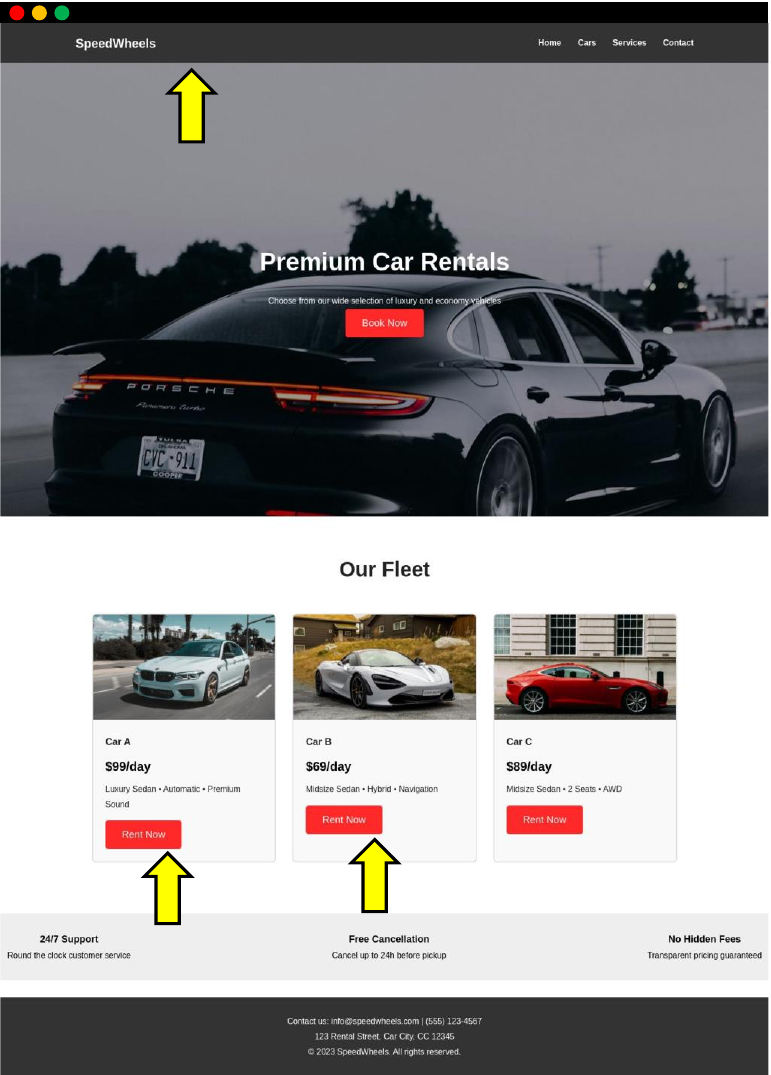}}
        \\
        
        \frame{\includegraphics[width=0.233\linewidth]{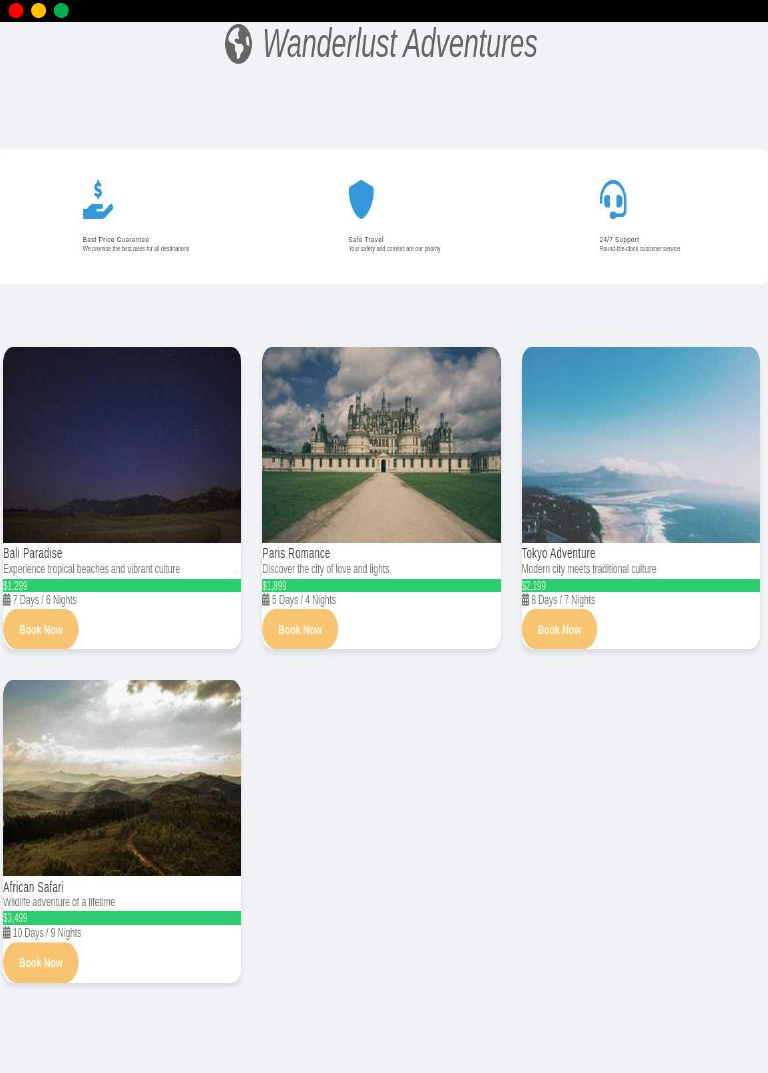}} &
        \frame{\includegraphics[width=0.233\linewidth]{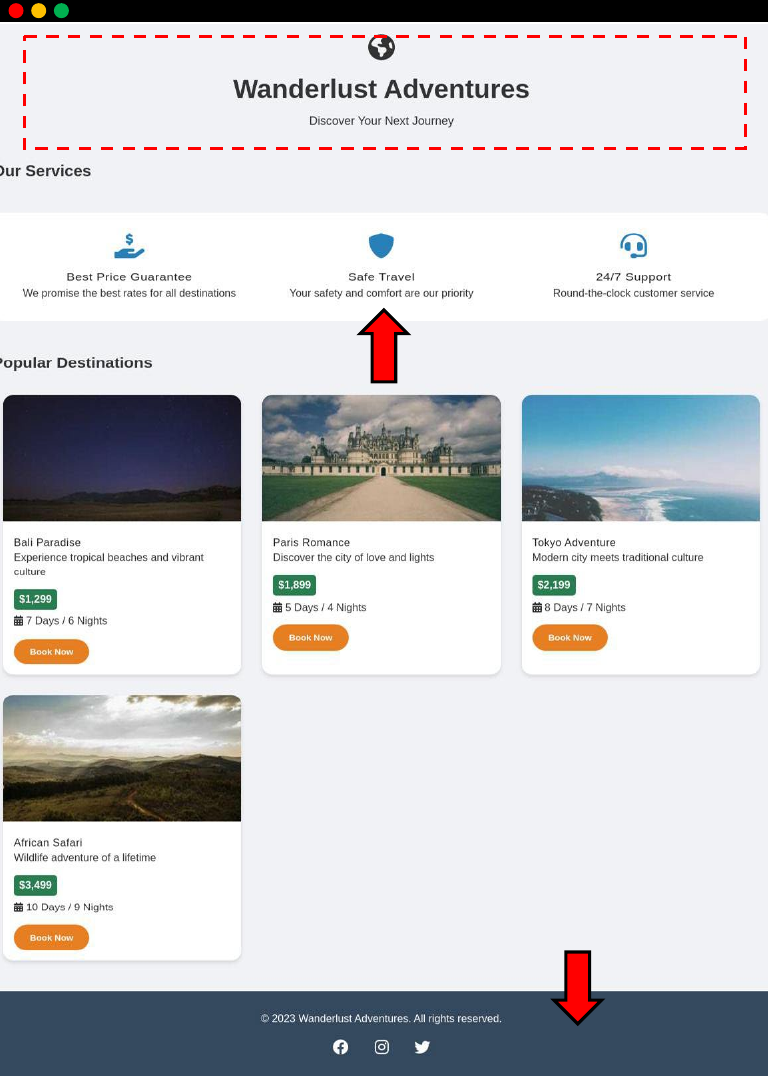}} &
        \frame{\includegraphics[width=0.233\linewidth]{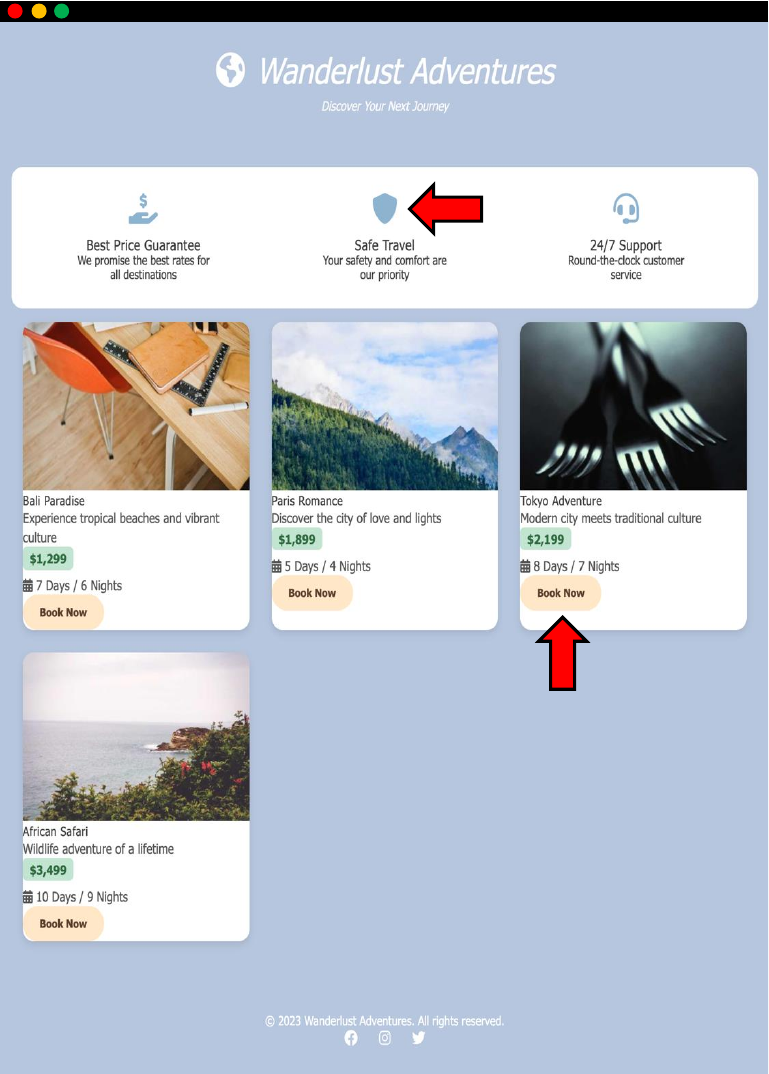}} &
        \frame{\includegraphics[width=0.233\linewidth]{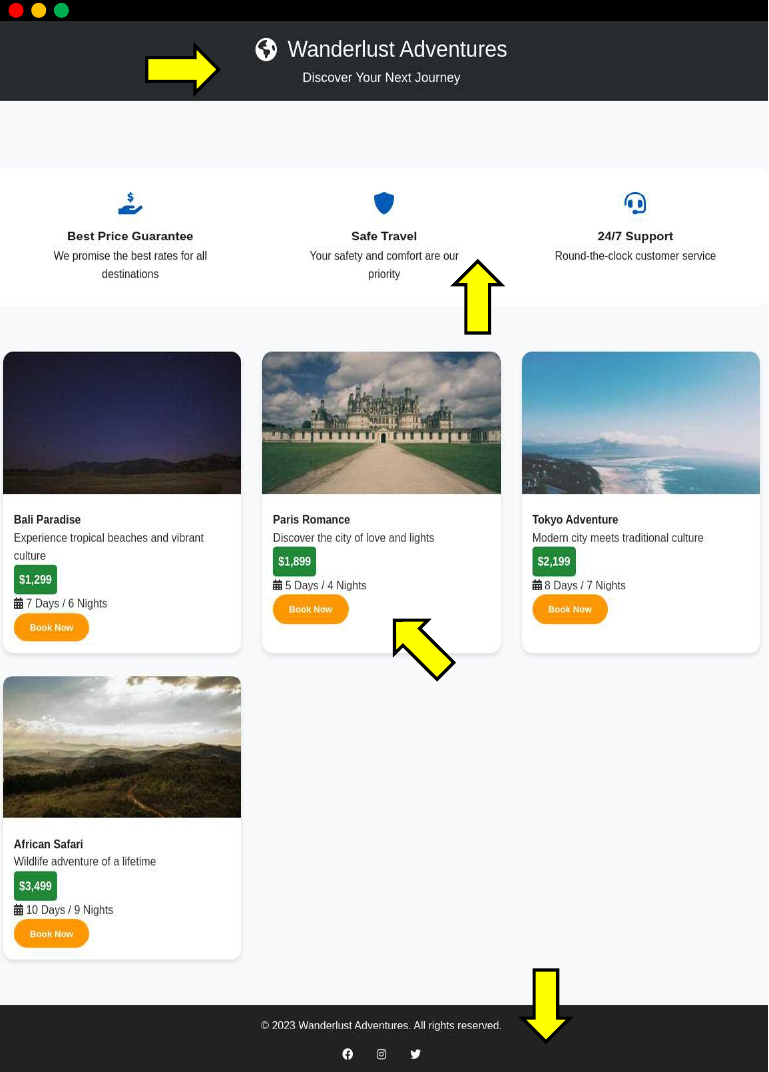}}
        \\        
    \end{tabular}
    \vspace{-0.3cm}
    \caption{Qualitative results with Ours, Claude 3.5 Sonnet~\cite{claude35sonnet}, and GPT-5~\cite{GPT5}. {Yellow} arrows highlight the fixed components, and {red} arrows show inaccessible parts.
    }
    \label{fig:qualitative_analysis}
    \vspace{-0.5cm}
\end{figure*}

\subsection{Qualitative Results \& Perceptual Study}\label{supp:user_study}

\myparagraph{Qualitative results.} \figref{fig:qualitative_analysis} presents qualitative comparisons of \textit{in-the-wild} websites to
show our model's ability to generalize beyond the \name\ dataset.
In the first example, our model corrects the header text contrast, adjusts the background image overlay, fixes button colors, and improves footer legibility, all while preserving the original layout.
Claude~3.5 only partially addresses the contrast issues: the hero text and primary navigation links remain inaccessible.
GPT-5 attempts a more aggressive fix but misaligns the header over the background image and introduces new low-contrast button colors. %

In the second example, our model fixes contrast violations in the header, icon labels, call-to-action buttons, and footer while keeping the page structure intact.
Claude~3.5 leaves the header and footer with insufficient contrast.
GPT-5 restructures the page significantly, dropping destination cards and repositioning the footer, consistent with its low structural accuracy in~\tabref{tab:main_table}.
Overall, our model fixes contrast violations while preserving the visual layout and content of the original website.

\myparagraph{Perceptual study design.}
We conducted a perceptual study to assess how well the modified HTML files preserve both the website's content and visual style. We compare against Claude~3.5 and GPT-5. Each participant is shown the reference webpage along with outputs from our method and two baselines, Claude 3.5 and GPT-5. Participants were asked, ``Which website (A, B, or C) best maintains the content and visual style of the reference image? If tied, select the one with the better visual design.''
Participants may select only one option for each question or leave the question empty. We collected 13 questions with responses from 30 participants.

\myparagraph{Statistical hypothesis setup.}
We analyze user preference data using a chi-squared test of independence to determine whether the distribution of selections differs significantly across the three methods. Let the three methods be denoted as $A$, $B$, and $C$ ($A$ is Ours, $B$ is Claude 3.5, and $C$ is GPT-5).
We first perform a chi-squared test of independence on the following hypothesis (More in~\secref{sec:test_stat}):
\bea
H_{0}^{\mathrm{global}} :& \; p_{A} = p_{B} = p_{C}, \\
H_{1}^{\mathrm{global}} :& \; \text{at least one }p_{i}\neq p_{j}.
\eea

If $H_{0}^{\mathrm{global}}$ is rejected at level $\alpha=0.05$, we proceed to post-hoc pairwise comparisons between each pair $(i,j)\in\{(A,B),(A,C)\}$. For each pair, we test
\bea
H_{0}^{\,i>j}: p_{i} = p_{j}, 
\qquad
H_{1}^{\,i>j}: p_{i} > p_{j}.
\eea
We apply the Bonferroni correction to control the family-wise error rate over the three comparisons. The adjusted significance threshold is $\alpha_{\mathrm{adj}} = \frac{0.05}{3} \approx 0.0167.$

\myparagraph{Results and statistical analysis.}
We aggregate selections over all 13 questions with 390 valid responses, with counts:
\bea
(n_A, n_B, n_C) = (195,\;76,\;119).
\eea
The chi-squared test of independence on the $3\times1$ contingency table yielded
\bea
\chi^2_{\mathrm{global}}(2,\,N=390) = 55.86,\quad p < 0.001.
\eea
Since $H_{0}^{\mathrm{global}}$ is rejected at $\alpha=0.05$, we perform post-hoc pairwise chi-squared tests with Bonferroni correction. The one-tailed comparisons for our method $A$ against $B$ and $C$ yield:
\bea
\chi^2_{A>B}(1,\,N_{A,B}=271) = 26.13, \quad p_{A>B} < 10^{-6},\\
\chi^2_{A>C}(1,\,N_{A,C}=314) = 9.20, \quad p_{A>C} < 0.002.
\eea
Both $p_{A>B}$ and $p_{A>C}$ are below $\alpha_{\mathrm{adj}} = 0.0167$. Participants prefer our method over both Claude~3.5 and GPT-5. Indicating that our method better preserves the website's content and visual style.

\section{Conclusion}
In this work, we formulate the task of improving website accessibility as an image-conditioned program synthesis problem. We construct WebAccessVL, a dataset of 1,500 paired HTML files manually modified to comply with WCAG2. We propose training VLMs in a violation-conditioned manner, where the model conditions on parsed violation reports describing specific accessibility issues detected in the input HTML. This conditioning enables the model to focus on fixing the violations in the report. Empirical results show that our method reduces violations from {5.34} to {0.211} per website with Gemma 3 and {0.244} with Llama 3.2 Vision, a {96.0\%} reduction from raw data. Compared to GPT-5 (1.68 violations), our approach achieves an {87\%} reduction while maintaining design fidelity, with 90\% structural accuracy vs. GPT-5's 0.5\%. Through systematic comparisons, we demonstrate that VLMs significantly outperform text-only LLMs, with fine-tuned VLMs achieving 60--77\% fewer violations than their LLM counterparts by leveraging visual understanding of rendered pages. This work represents an important step toward automating web accessibility. To support further research, we will release both the code and the dataset.

\clearpage
{
    \small
    \bibliographystyle{splncs04}
    \bibliography{ref}
}

\clearpage
\onecolumn 

{\bf\noindent \Large Appendix}\\

\setcounter{section}{0}
\renewcommand{\theHsection}{A\arabic{section}}
\renewcommand{\thesection}{A\arabic{section}}
\renewcommand{\thetable}{A\arabic{table}}
\setcounter{table}{0}
\setcounter{figure}{0}
\renewcommand{\thetable}{A\arabic{table}}
\renewcommand\thefigure{A\arabic{figure}}
\renewcommand{\theHtable}{A.Tab.\arabic{table}}%
\renewcommand{\theHfigure}{A.Abb.\arabic{figure}}%
\renewcommand\theequation{A\arabic{equation}}
\renewcommand{\theHequation}{A.Abb.\arabic{equation}}%

{\bf\noindent The appendix is organized as follows:}
\begin{itemize}%
    \item We describe the details of each accessibility violation in~\secref{supp_sec:violation_details}.
    \item We provide additional qualitative results in~\secref{supp:additional_qual}. Also see the attached HTML files in the supplementary materials.
    \item We conduct ablation studies on violation conditioning strategies and negative guidance in~\secref{sec:ablation_study}.
    \item We analyze the computational efficiency of our checker-in-the-loop approach in~\secref{sec:loop_efficiency}.
    \item Implementation, experimentation details, and additional prompt experiments are provided in~\secref{supp:implementation_details}. We will release the code and dataset.
    \item We explain the test statistic for the user study in~\secref{sec:test_stat}.
    \item We report the training and inference times of our violation-conditioned VLM in~\secref{sec:train_inf_time}.
    \item We document how we annotated websites to improve WCAG2 compliance for the WebAccessVL dataset in~\secref{sec:website_correction}.
\end{itemize}

\section{Dataset Violation Details}
\label{supp_sec:violation_details}
\noindent{\textbf{Vision violations.}}
Below, we describe the top-five vision violations:\\
{\null\quad \ding{172}} \texttt{Text contrast sufficient} violation means that the ratio of text contrast and background is less than a 3 to 1 contrast ratio. This violation requires visual understanding to fix the color contrast while maintaining the website's design consistency. Naively using a white background and black text would resolve this violation, but it would also lead to large changes in the website design.\\ 
{\null\quad \ding{173}} \texttt{Img alt valid} violation is triggered when there is a missing alternative text. Each visual representation, such as a product image, logo, or any image, is required to have a short description of the image. For example, the checker verifies that the \texttt{<img>} tag has an \texttt{alt} attribute or not in the HTML. Vision information is needed as the description needs to match the image content.\\
{\null\quad \ding{174}} The checker flags a \texttt{style color misuse} violation when there are only color differences to mark required fields, \ie, users who cannot perceive the color will not be able to perform further actions. 
This violation can be fixed using a visual indication such as the asterisk (*) in the required field. Note that the color needs to be chosen carefully; an inappropriate color could recursively trigger color contrast violations.\\
{\null\quad \ding{175}} \texttt{caption track exists} violation occurs when there is no textual information on a given video content. This violation can be solved by providing captions containing textual information on the video content.\\ 
{\null\quad \ding{176}} \texttt{svg graphics labeled} violation is similar to a missing alt text, but it differs in terms of format.  SVG can have multiple layers of components, and each component should have its corresponding label, \ie, labeling the meaning for all the components. This additional information can be provided using an accessible specific tag to assistive technologies using \texttt{aria-label}; ARIA stands for Accessible Rich Internet Applications.

In summary, these vision violations need some degree of reasoning over the visual content of the website and would be difficult to address using only language information.

\noindent{\textbf{Language violations.}}
We now describe the top-five language violations:\\
{\null\quad   \ding{182}} \texttt{aria content in landmark} violation is when there is no role or subsection specified. Without having the role, screen readers or other assistive technologies cannot understand the elements on the web page. A textual understanding of content is required to assign a proper role from the eight possible roles, %
using the \texttt{role} tag.\\
{\null\quad \ding{183}} \texttt{html lang exists} violation triggers when there is no specified type of language on the website. It is necessary to support assistive technology such as speech synthesizers to adapt the pronunciation to the specific language.\\ %
{\null\quad  \ding{184}} \texttt{page title exists} violation is marked if there is no \texttt{head} element and \texttt{title} element in HTML codes. A page title is important because it is the first part that assistive technology accesses the website and tells the users so that they can confirm they are accessing the right website.\\ %
{\null\quad  \ding{185}} \texttt{skip main exists} violation happens when \texttt{main} tag is missing. Without the tag, users with assistive technologies cannot directly go to the \texttt{main} elements, and users need to go through all unnecessary content, such as banners or navigation elements.\\ %
{\null\quad \ding{186}} \texttt{text block heading} violation is when there are no headings, \ie, the website has dense blocks of text. With proper headings, they allow users with assistive technology to go through blocks by accessing a summary of contents and quickly navigating to other sections. 

Overall, these language violations are mainly related to the HTML structuring and summarizing of the website's text content. Addressing these violations will not involve the usage of vision information.

\section{Additional Qualitative Results}
\label{supp:additional_qual}
\begin{figure*}[ht]
\centering
\begin{tabular}{ccc>{\columncolor{mygray}}c}
 Input & Claude3.5 \\ 
         \frame{\includegraphics[width=0.47\linewidth]{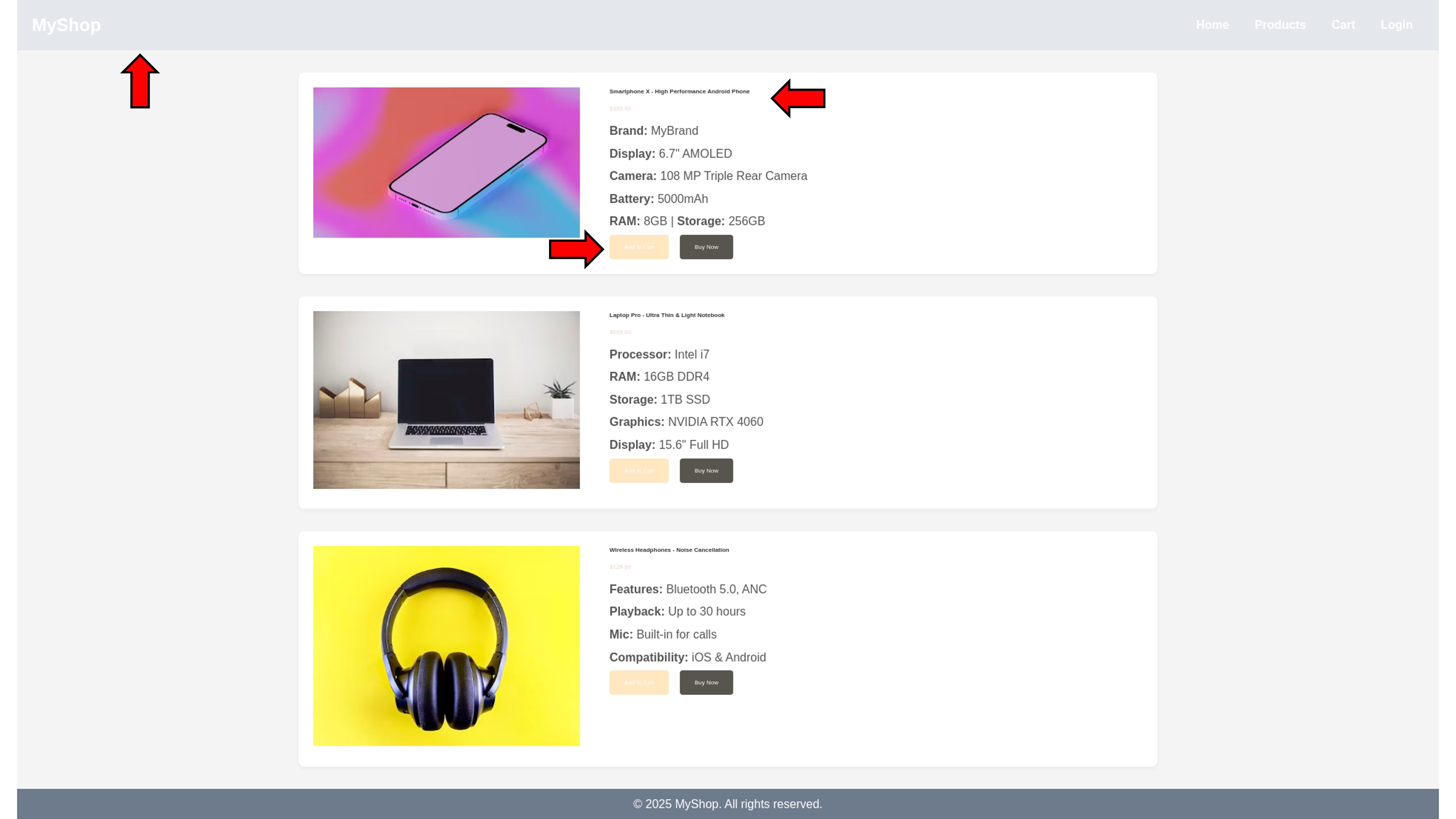}} &
        \frame{\includegraphics[width=0.47\linewidth]{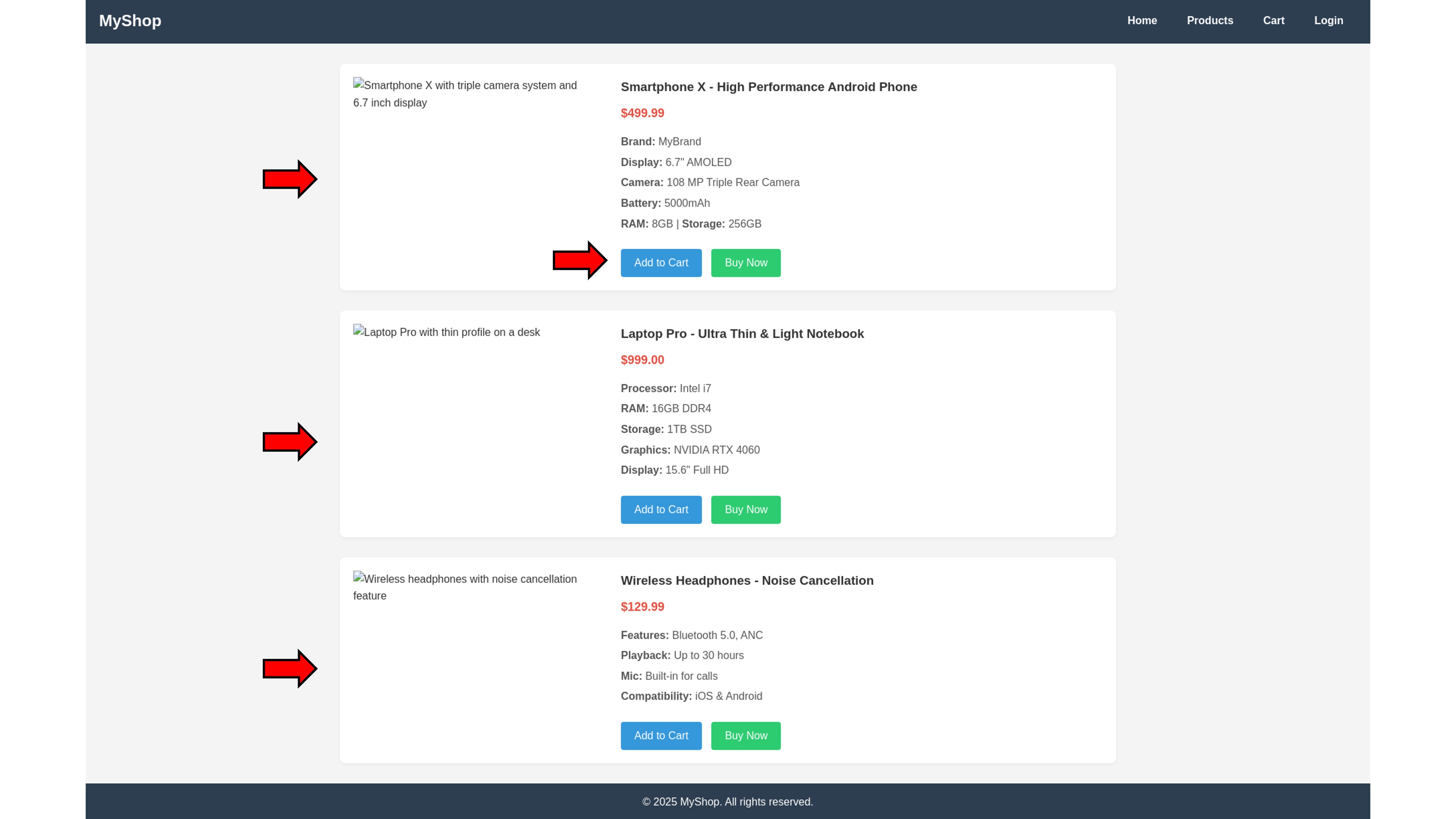}} \\
        GPT-5 & \bf Ours\\
        \frame{\includegraphics[width=0.47\linewidth]{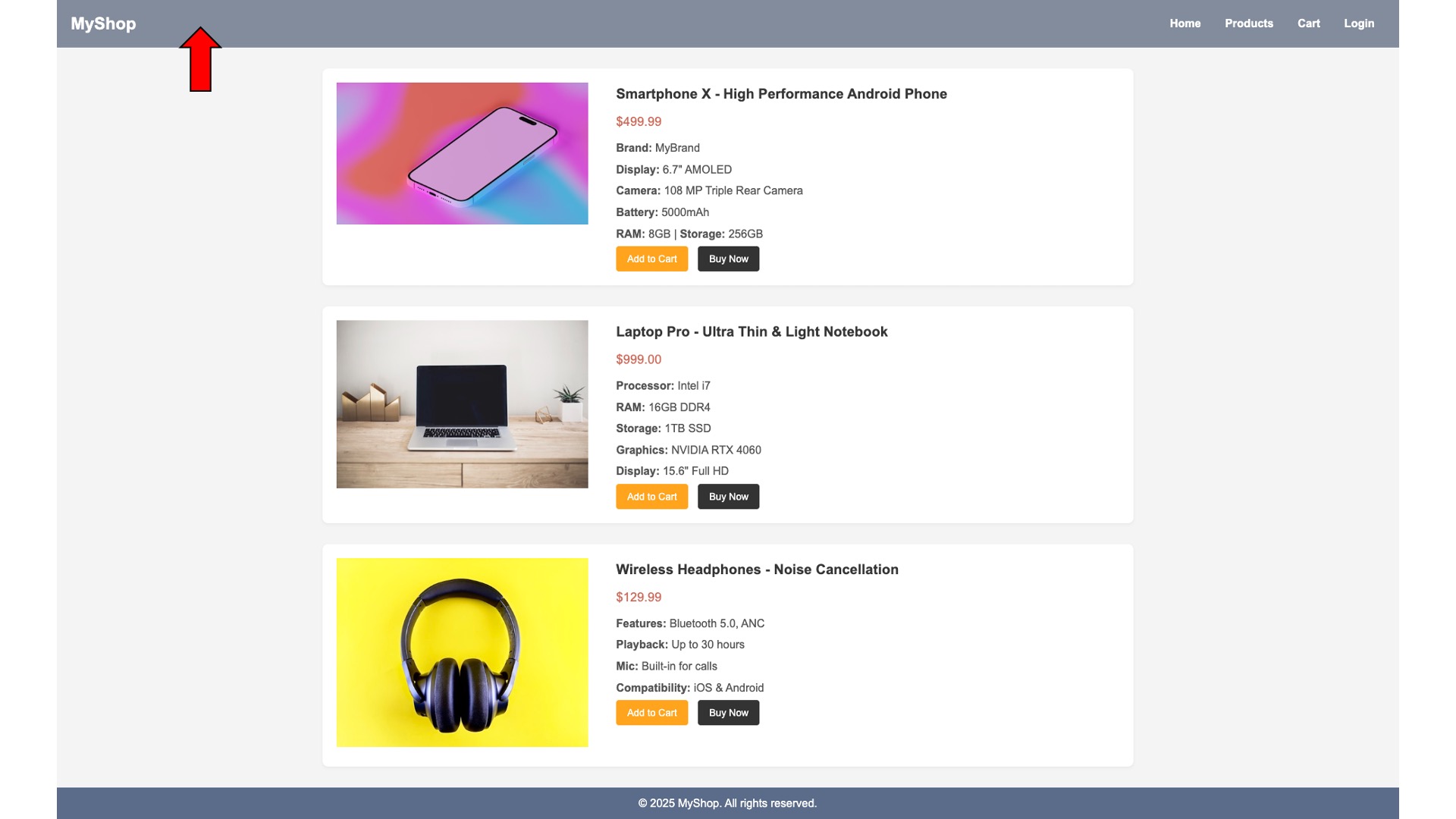}}  &      \frame{\includegraphics[width=0.47\linewidth]{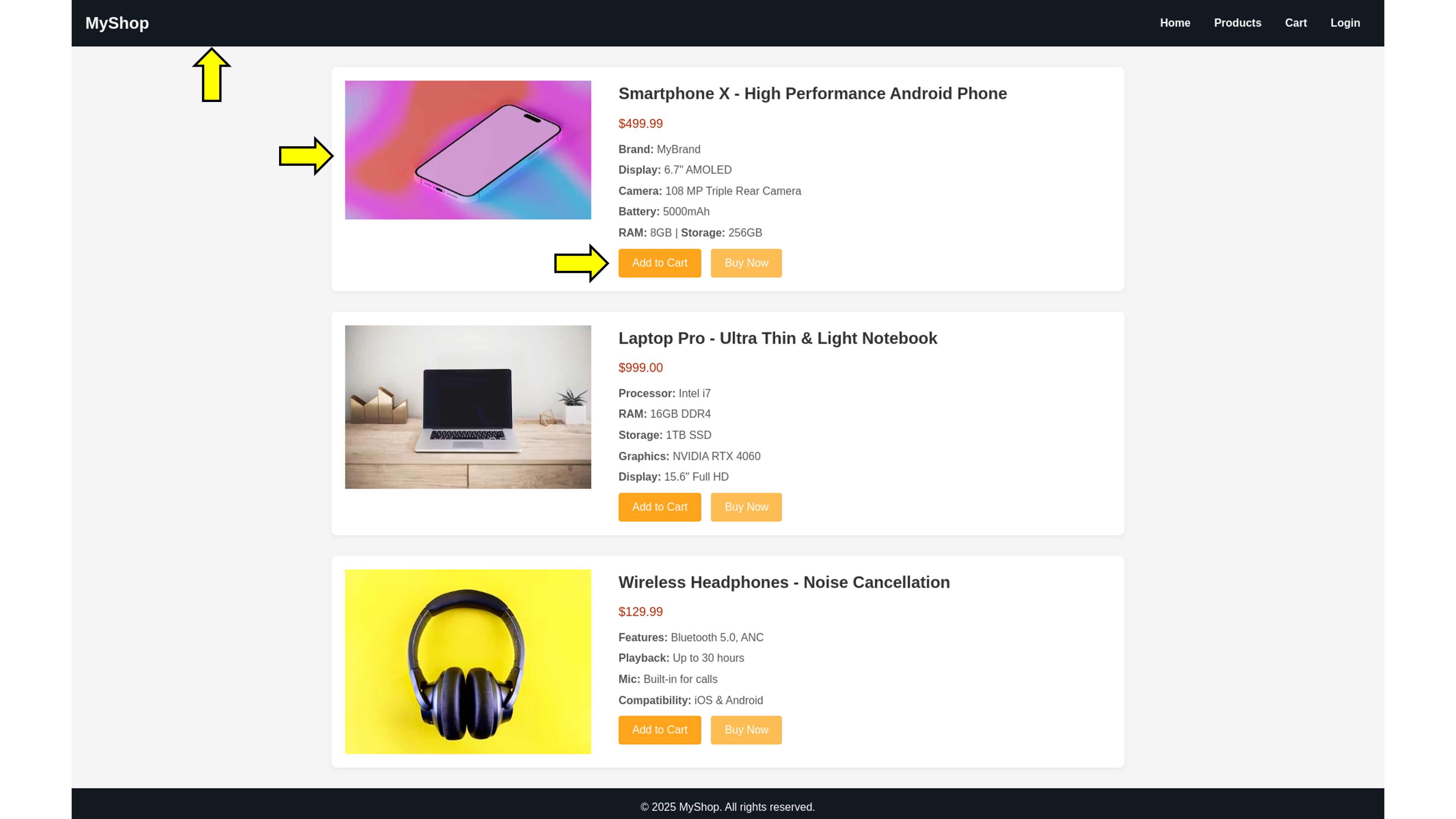}}\\
\end{tabular}        
    \caption{Yellow arrows highlight the fixed components, and red arrows show inaccessible parts. Ours modifies the contrast between the header, a button, and the descriptions in each card, while Claude 3.5 fails to put valid image links. Moreover, GPT-5 could not correct the contrast of the header. Please zoom in for details.
    }
    \label{fig:appendix_q1}
\end{figure*}

\begin{figure*}[ht]
\centering
\begin{tabular}{ccc>{\columncolor{mygray}}c}
 Input & Claude3.5 \\ 
         \frame{\includegraphics[width=0.47\linewidth]{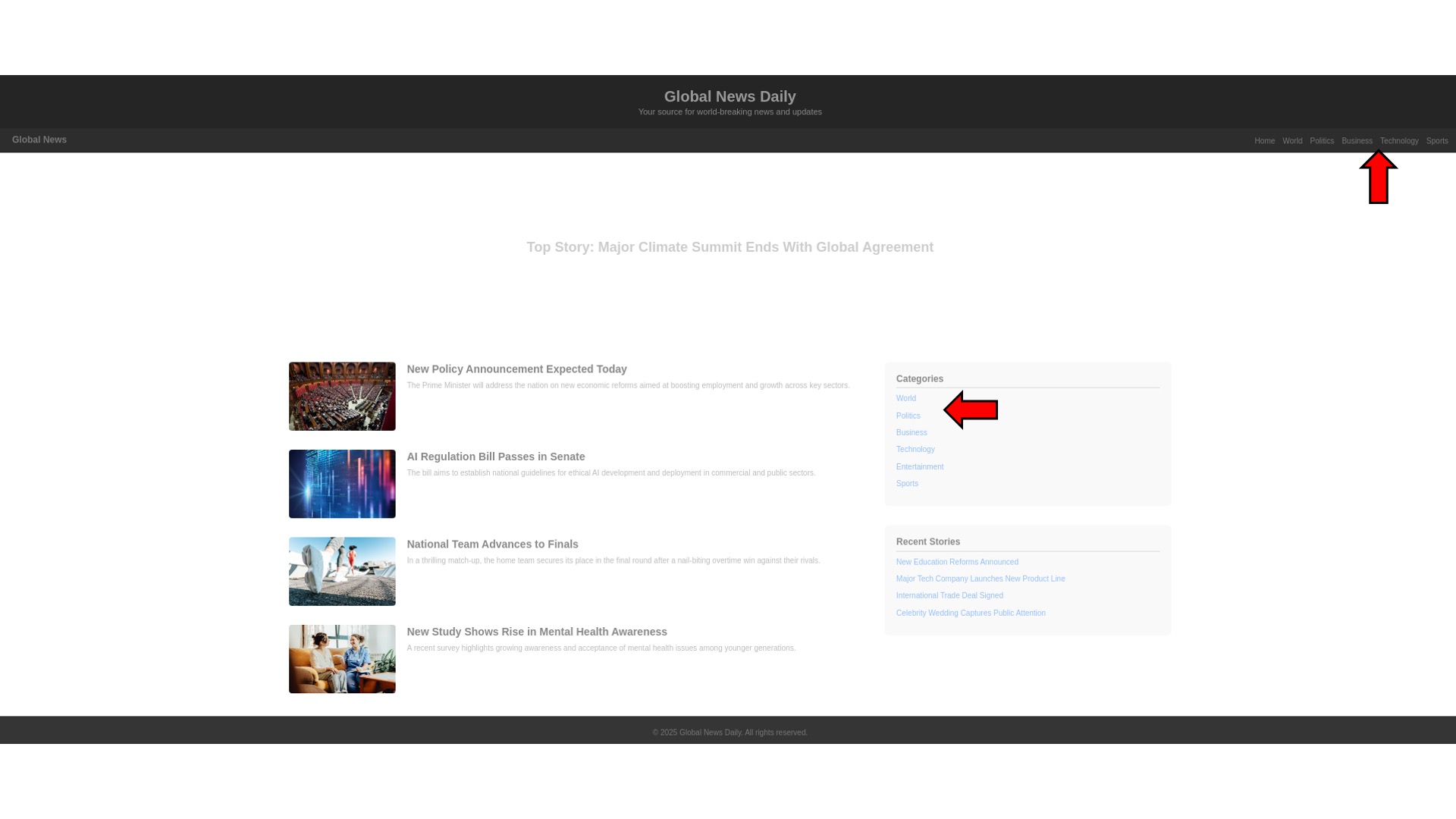}} &
        \frame{\includegraphics[width=0.47\linewidth]{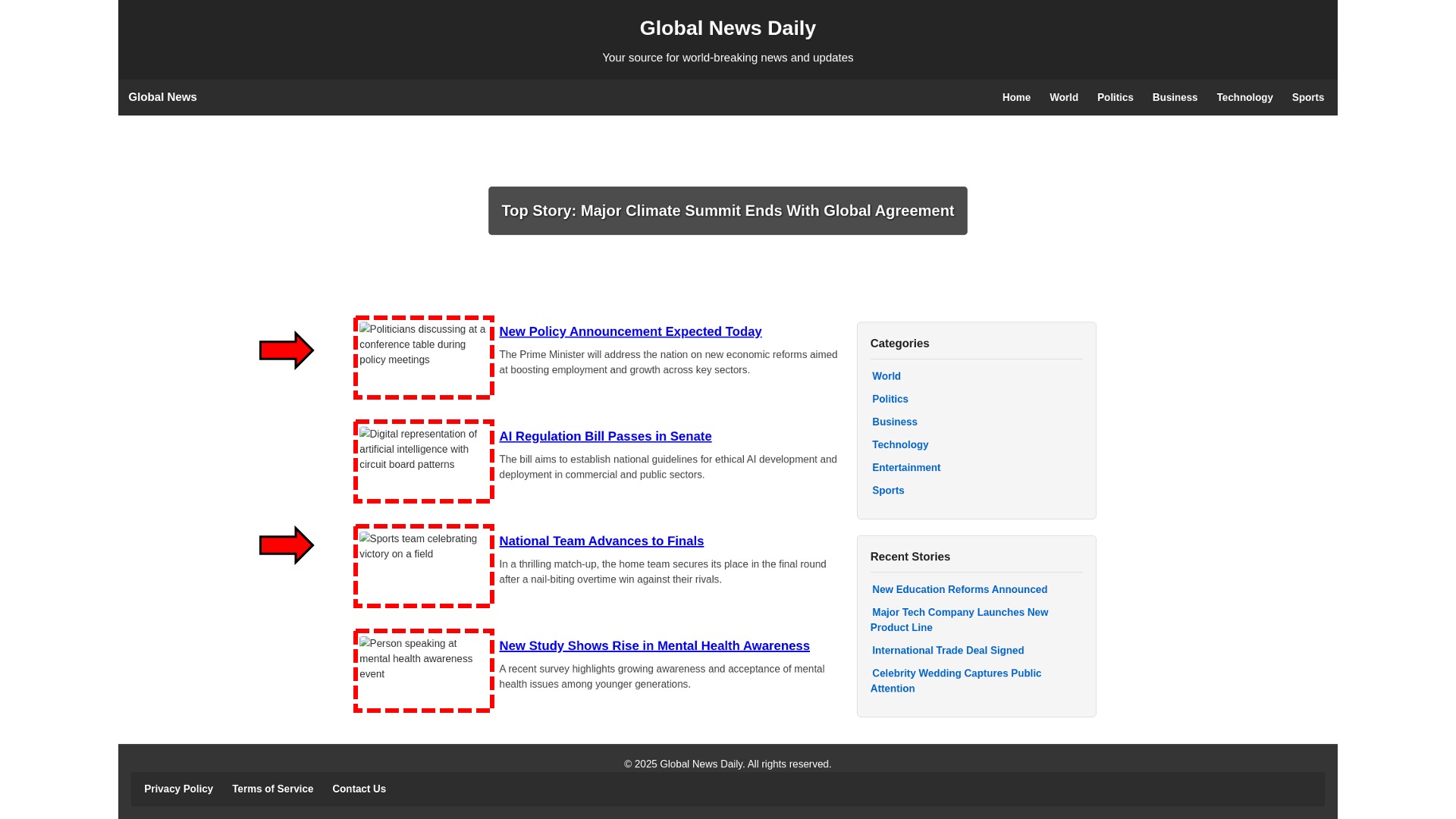}} \\
        GPT-5 & \bf Ours\\
        \frame{\includegraphics[width=0.47\linewidth]{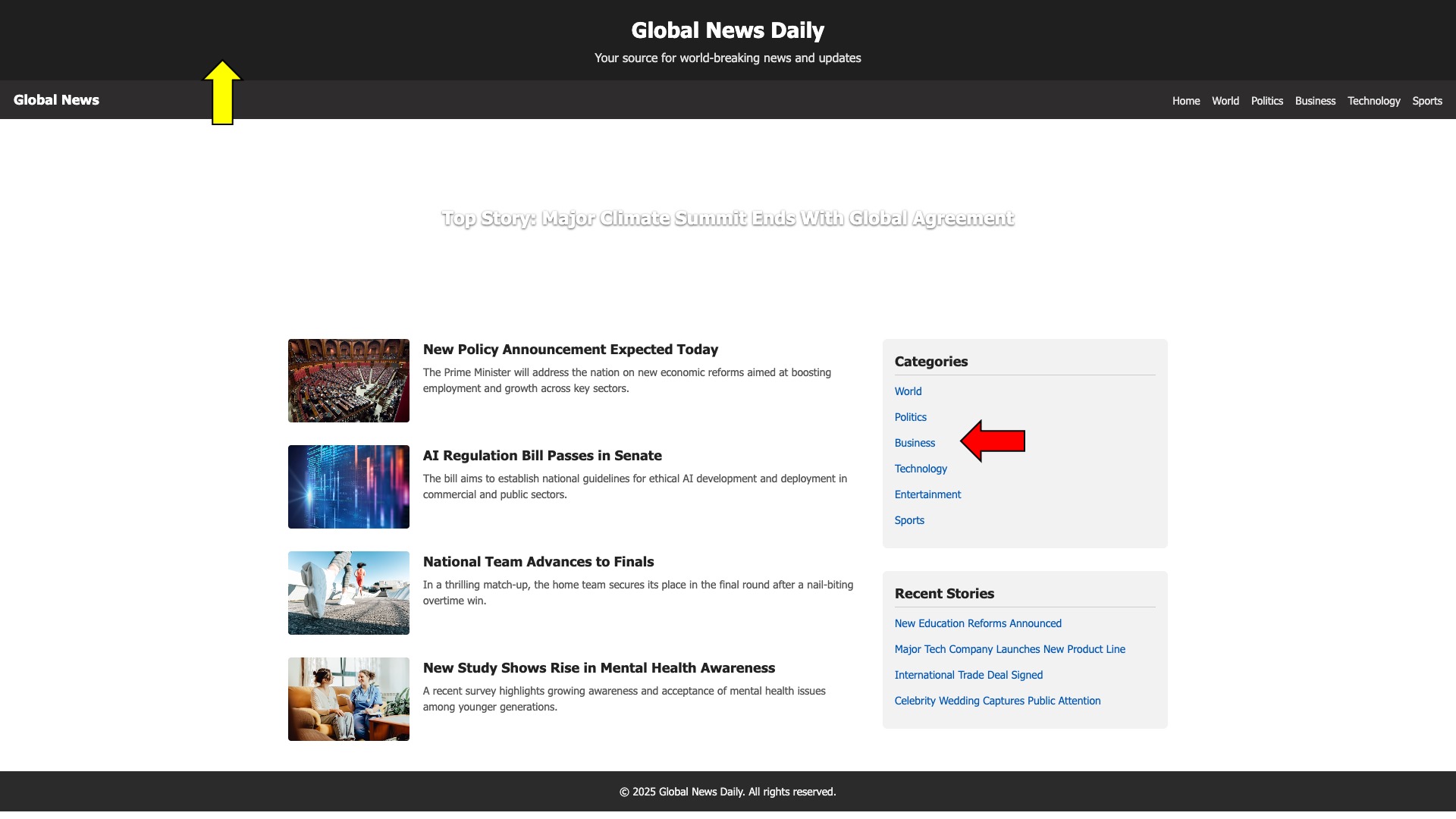}}  &      \frame{\includegraphics[width=0.47\linewidth]{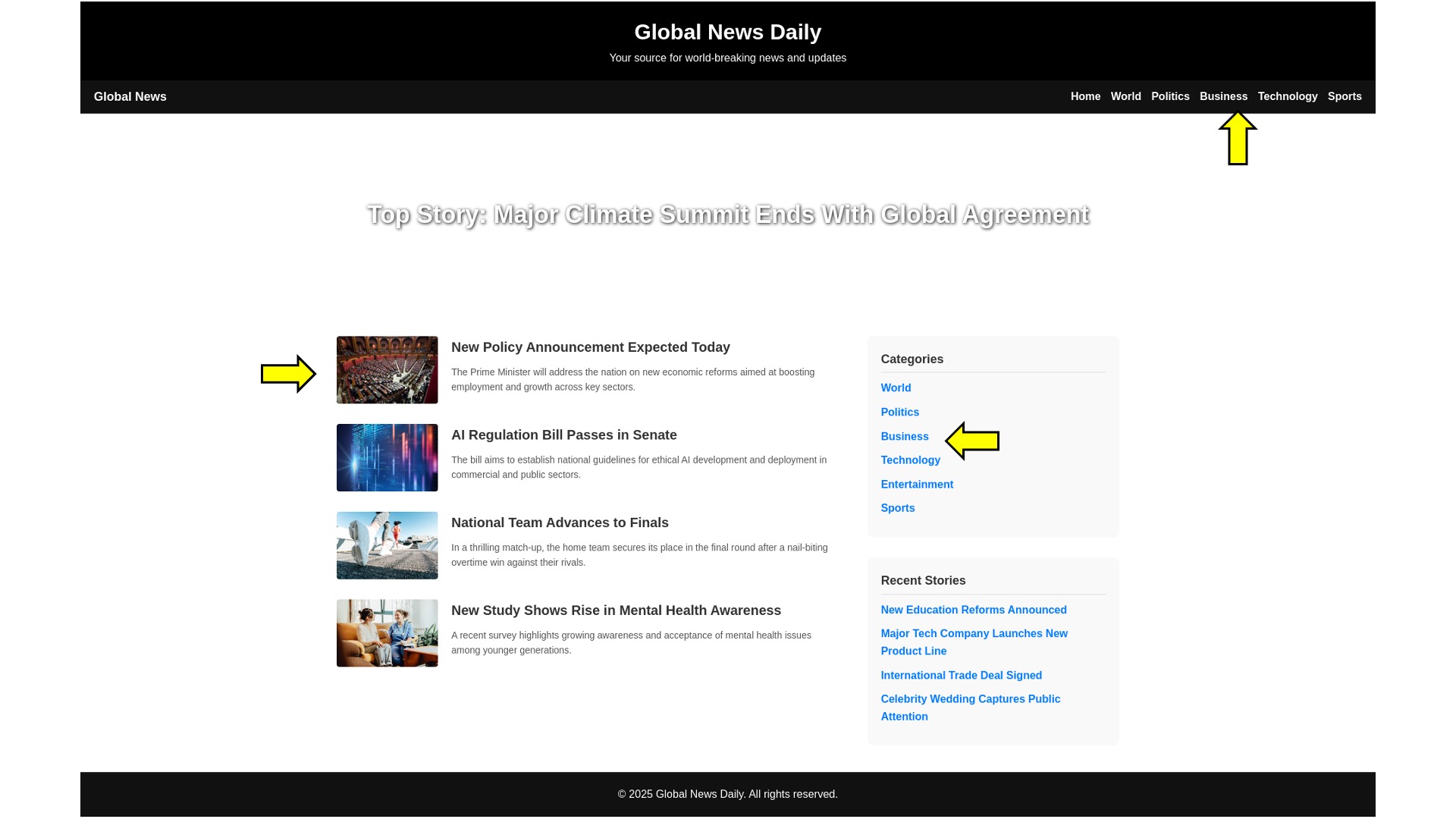}}\\
\end{tabular}        
    \caption{Yellow arrows highlight the fixed components, and red arrows show inaccessible parts. Ours corrects the text contrast in the navigation bar, descriptions in each image, and sidebars. Claude 3.5 does not have appropriate contrast on the title of each card and is missing the thumbnails. GPT-5 also has not completely fixed the text contrast in the left sidebar. Please zoom in for details.
    }
    \label{fig:appendix_q2}
\end{figure*}

\begin{figure*}[ht]
\centering
\begin{tabular}{ccc>{\columncolor{mygray}}c}
 Input & Claude3.5 \\ 
         \frame{\includegraphics[width=0.47\linewidth]{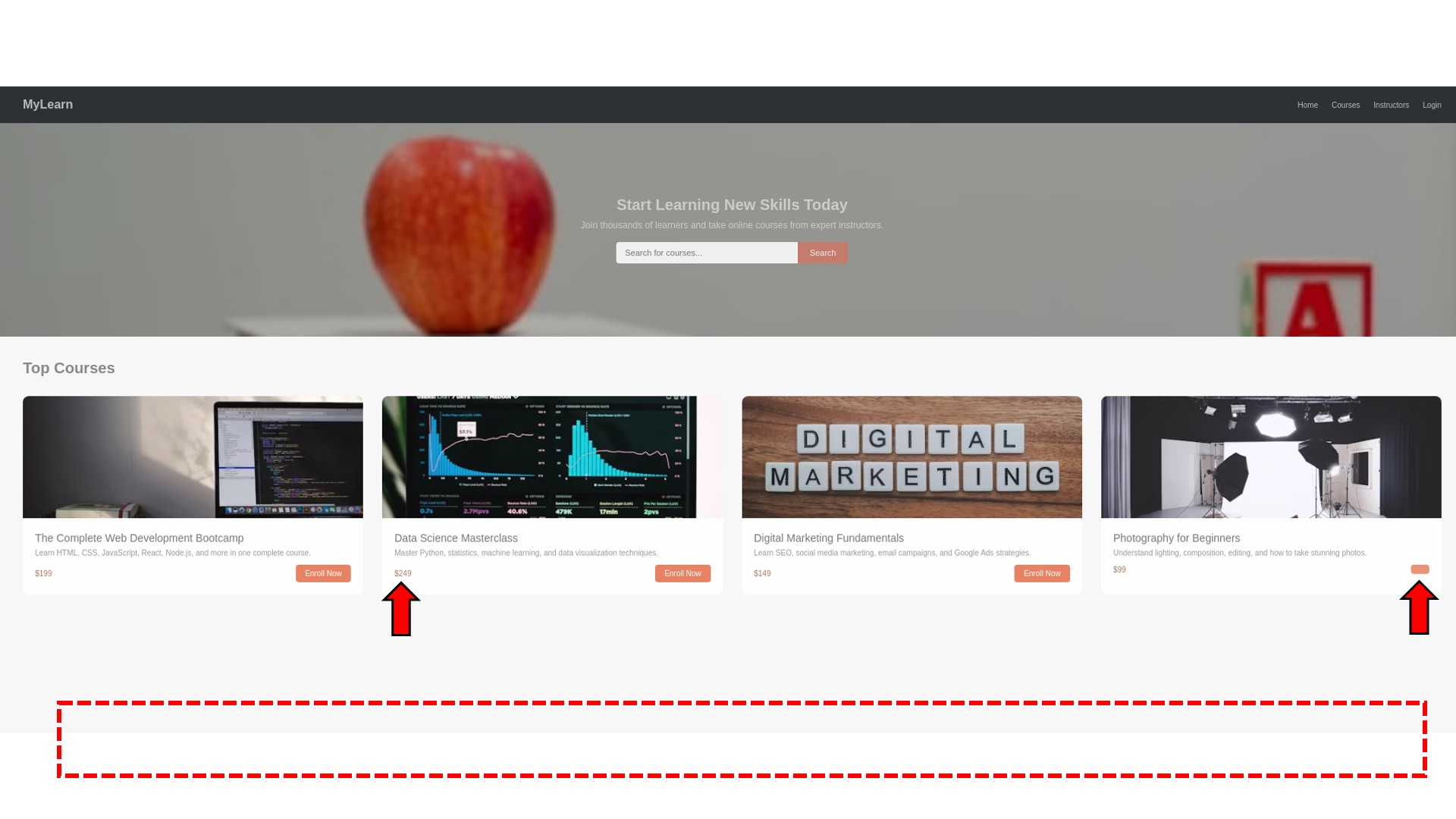}} &
        \frame{\includegraphics[width=0.47\linewidth]{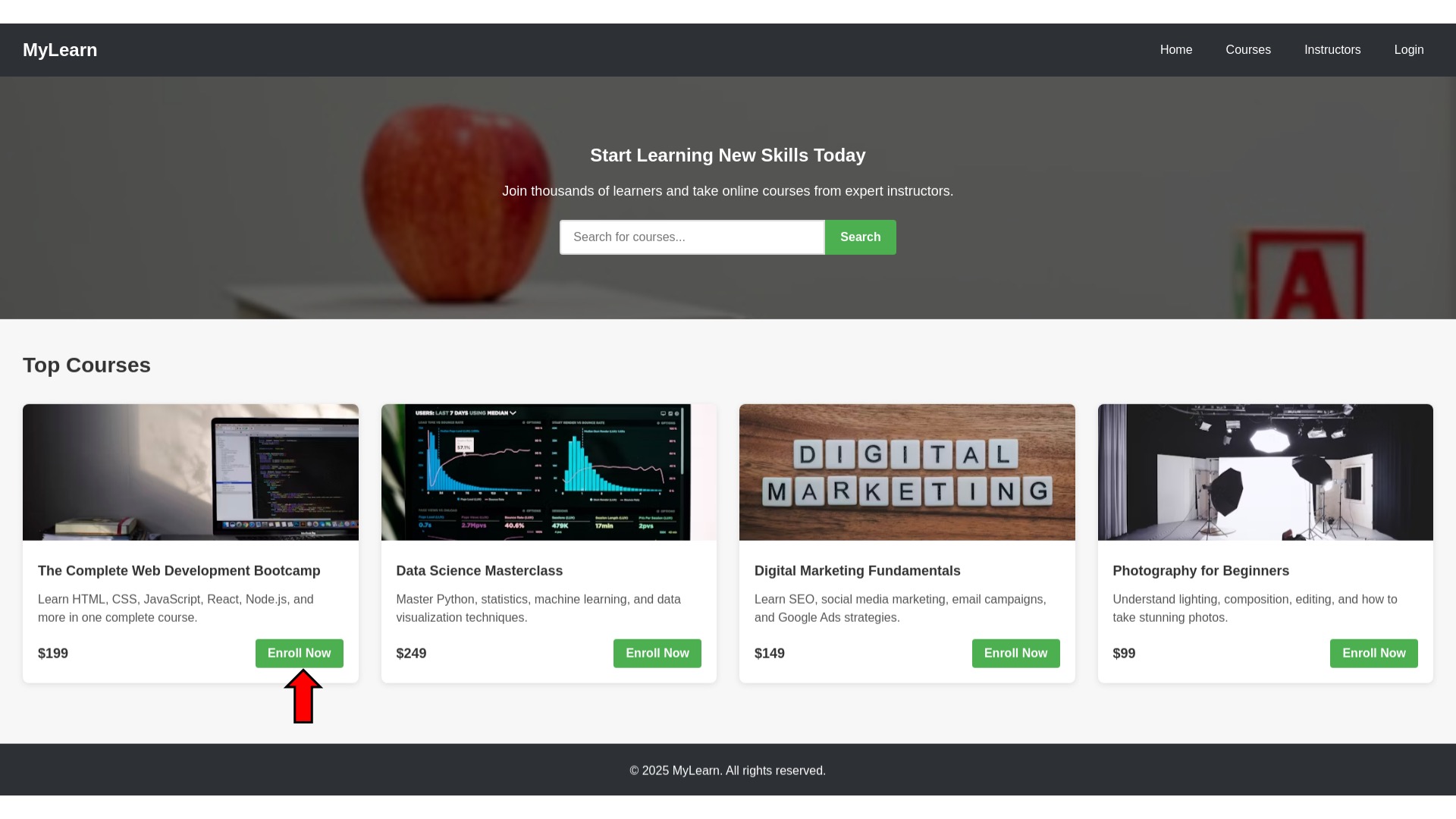}} \\
        GPT-5 & \bf Ours\\
        \frame{\includegraphics[width=0.47\linewidth]{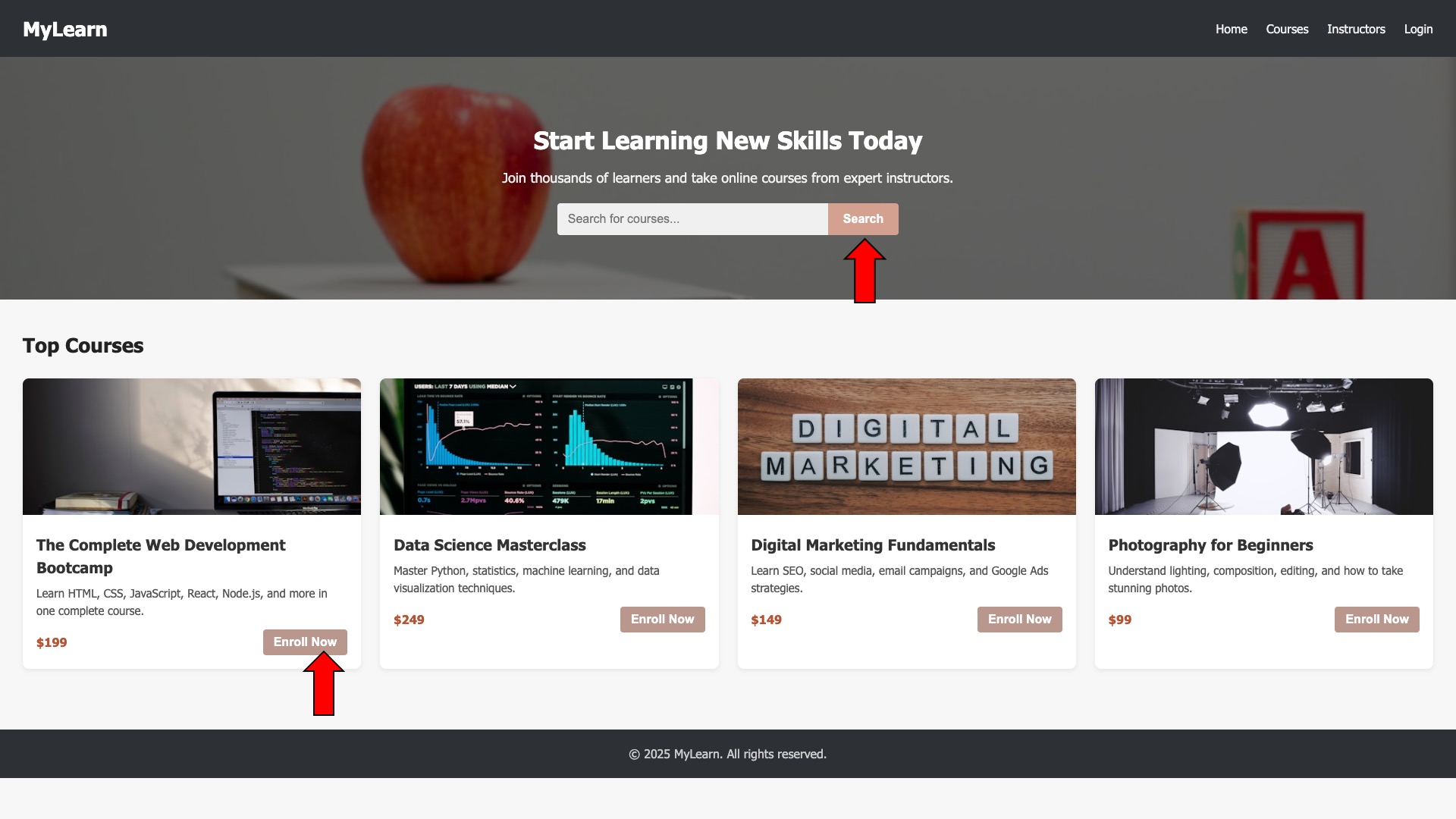}}  &      \frame{\includegraphics[width=0.47\linewidth]{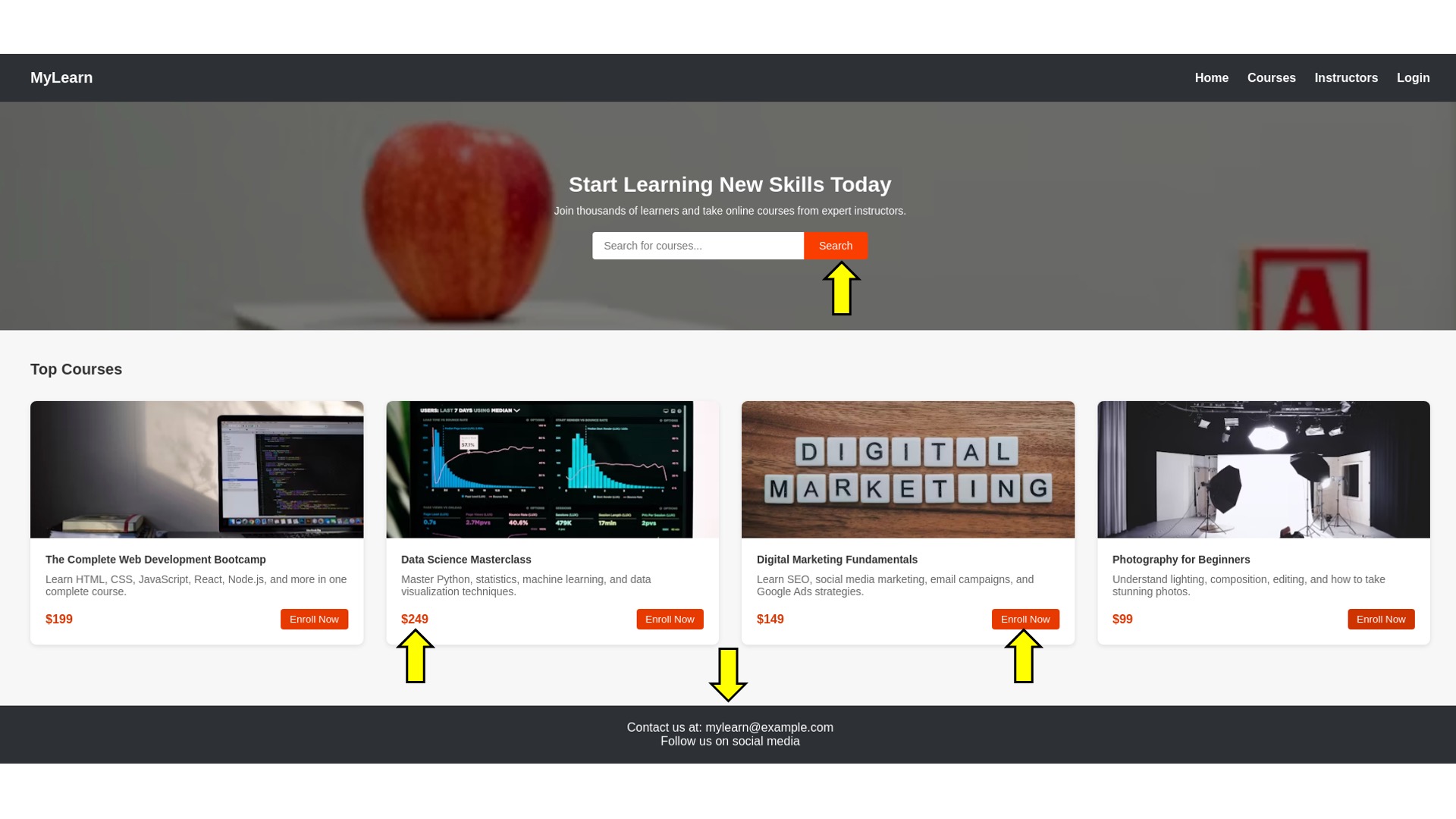}}\\
\end{tabular}        
    \caption{Yellow arrows highlight the fixed components, and red arrows show inaccessible parts. Ours successfully put an accessible contrast in buttons and the texts. Claude 3.5 and GPT-5 failed to put an appropriate color contrast on the buttons. Please zoom in for details.
    }
    \label{fig:appendix_q4}
\end{figure*}

\begin{figure*}[ht]
\centering
\begin{tabular}{ccc>{\columncolor{mygray}}c}
 Input & Claude3.5 \\ 
         \frame{\includegraphics[width=0.47\linewidth]{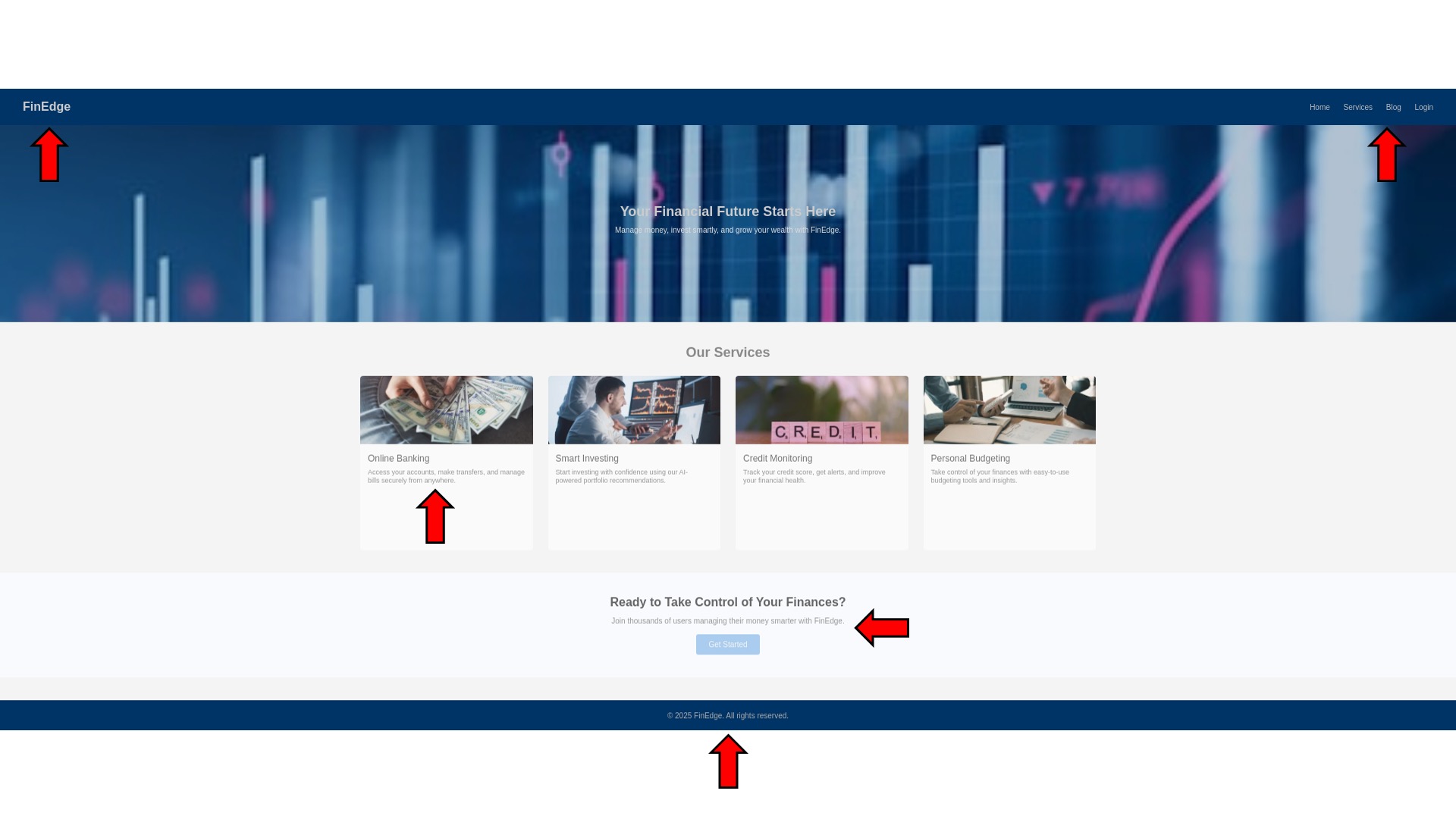}} &
        \frame{\includegraphics[width=0.47\linewidth]{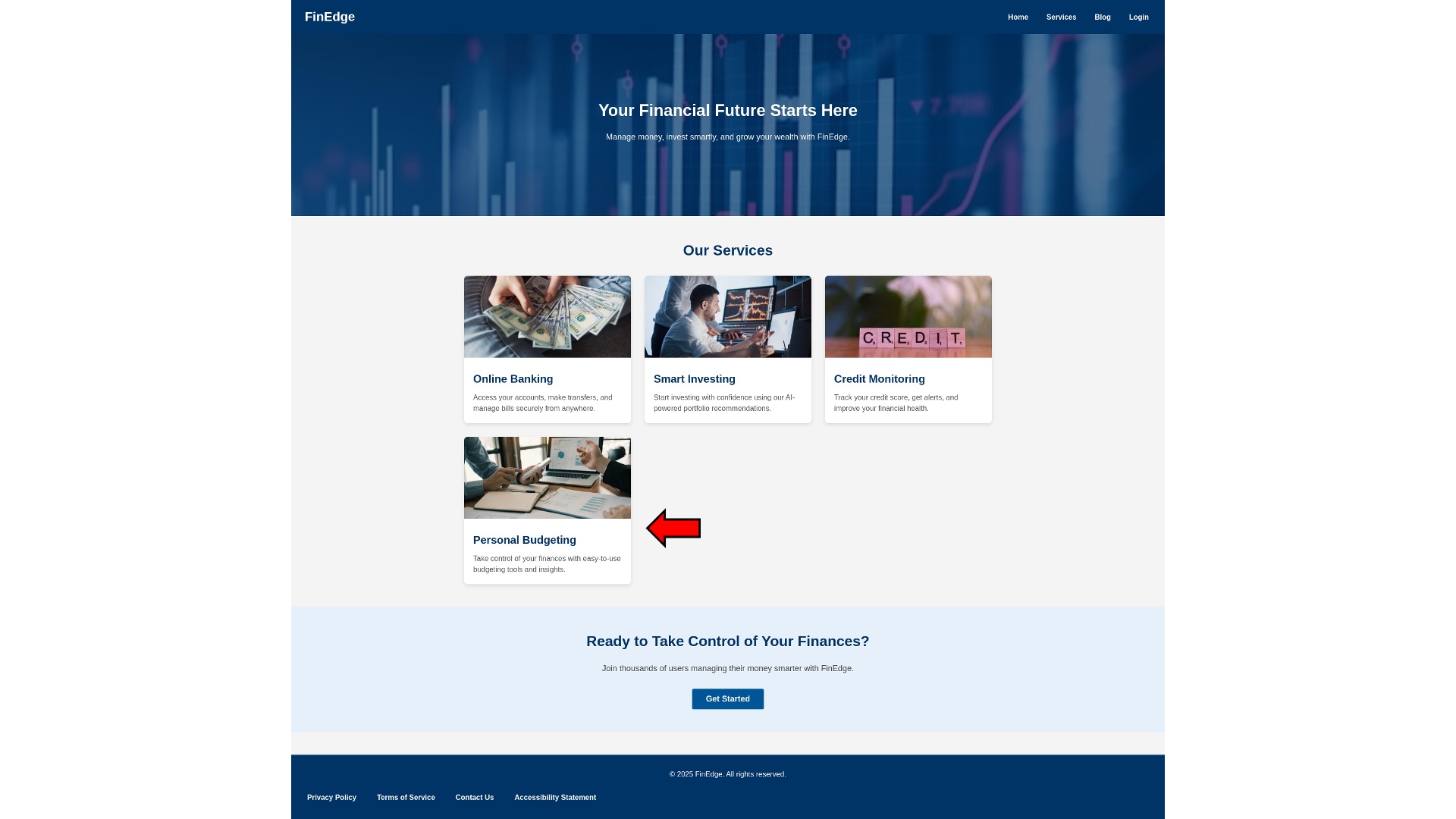}} \\
        GPT-5 & \bf Ours\\
        \frame{\includegraphics[width=0.47\linewidth]{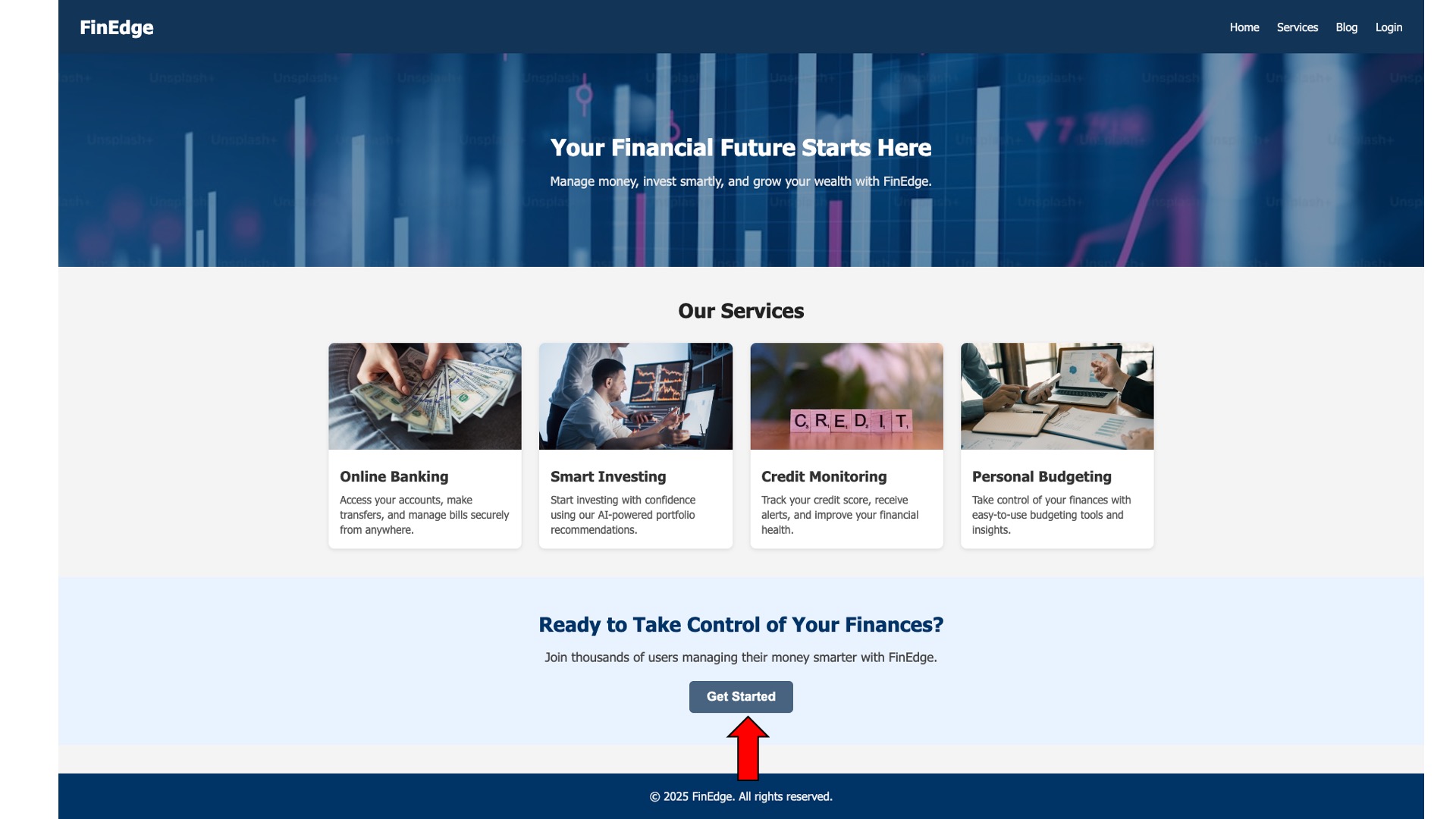}}  &      \frame{\includegraphics[width=0.47\linewidth]{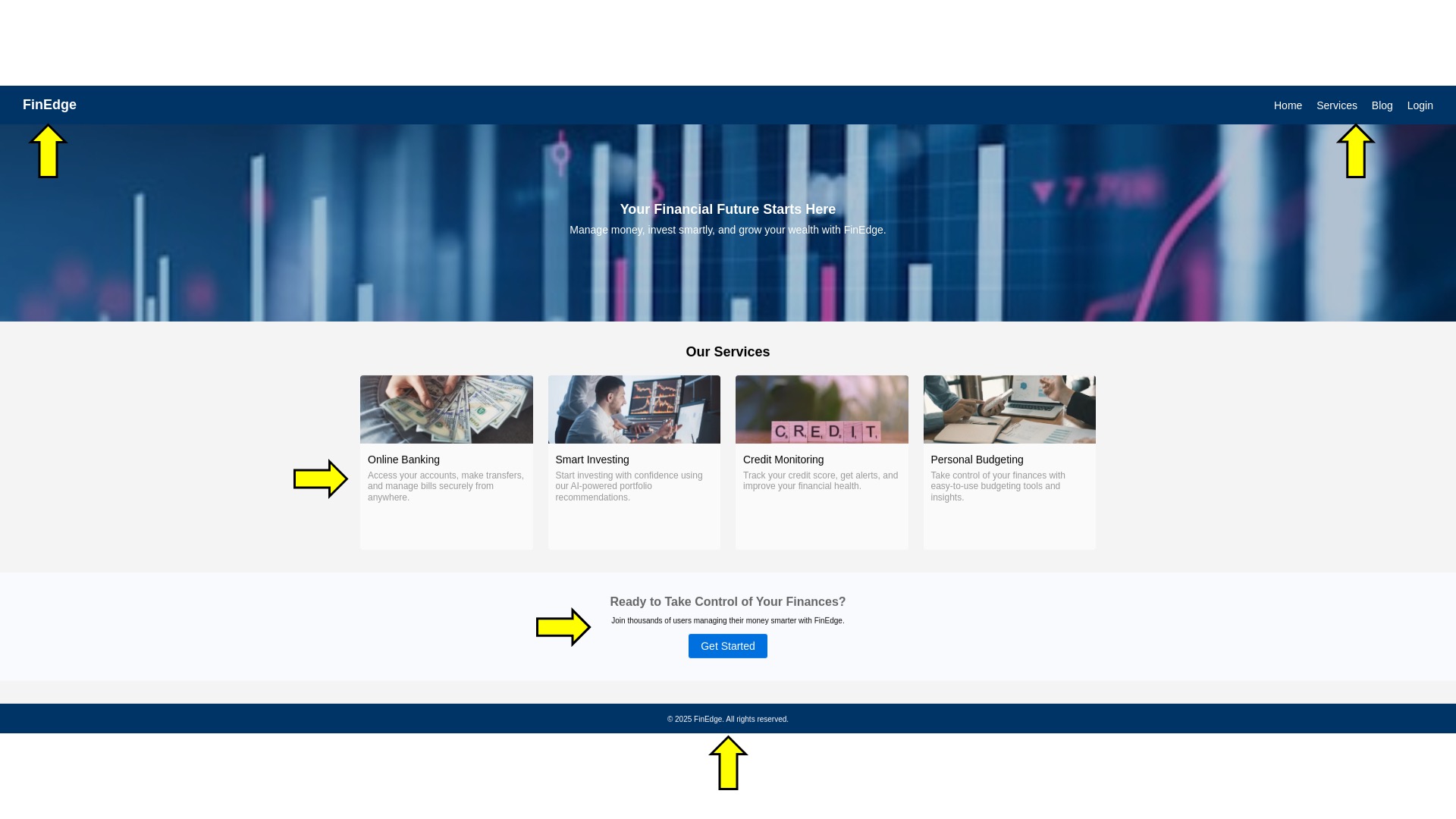}}\\
\end{tabular}        
    \caption{Yellow arrows highlight the fixed components, and red arrows show inaccessible parts. Ours shows that it generates an accessible contrast on the header, description, and footer. GPT-5 fails to put a consistent blue color in the button. Moreover, ours preserves the website style by putting all four images in a row, while Claude 3.5 could not preserve the original design. Please zoom in for details.
    }
    \label{fig:appendix_q5}
\end{figure*}

\begin{figure*}[ht]
\centering
\begin{tabular}{ccc>{\columncolor{mygray}}c}
 Input & Claude3.5 \\ 
         \frame{\includegraphics[width=0.47\linewidth]{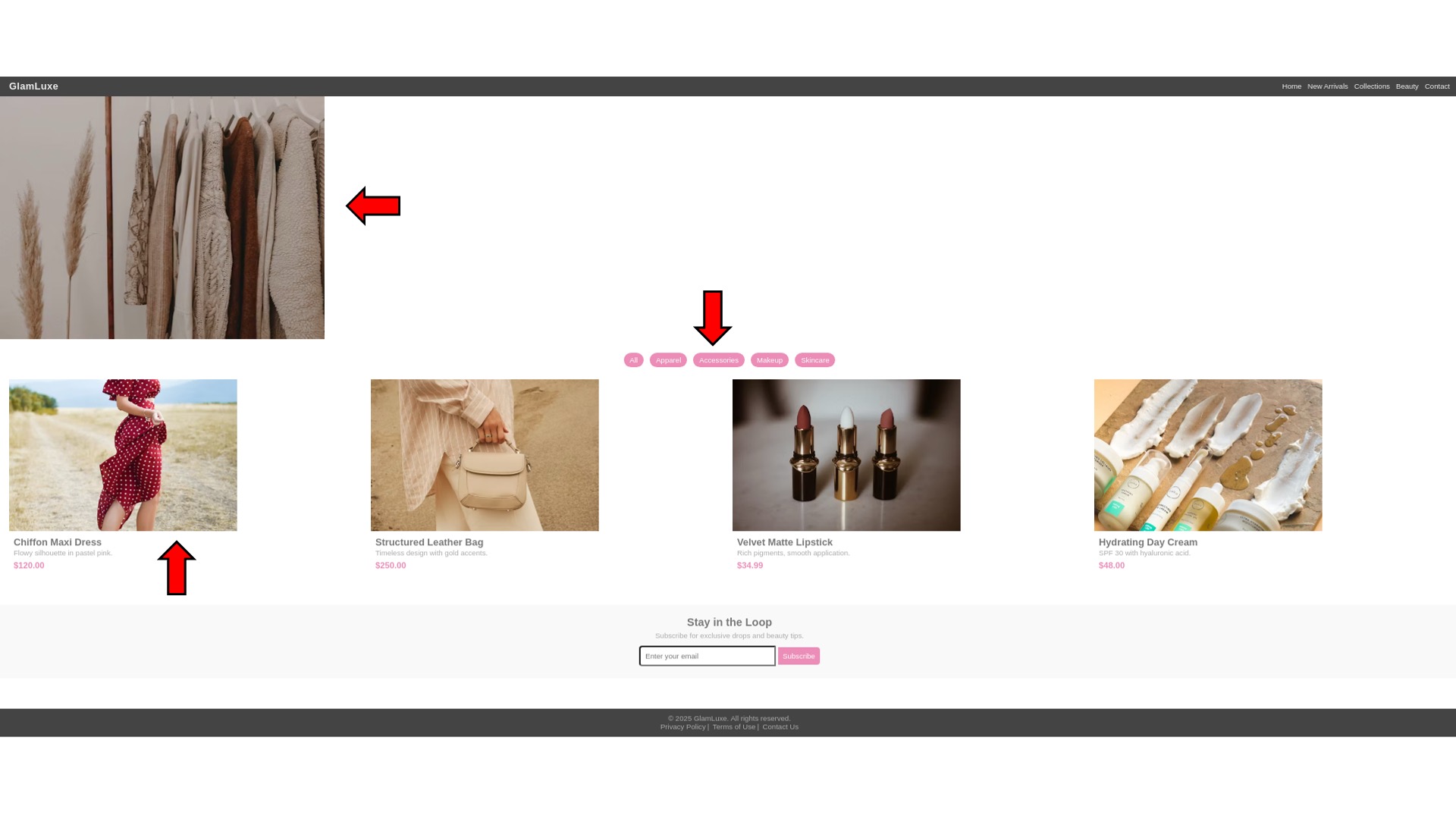}} &
        \frame{\includegraphics[width=0.47\linewidth]{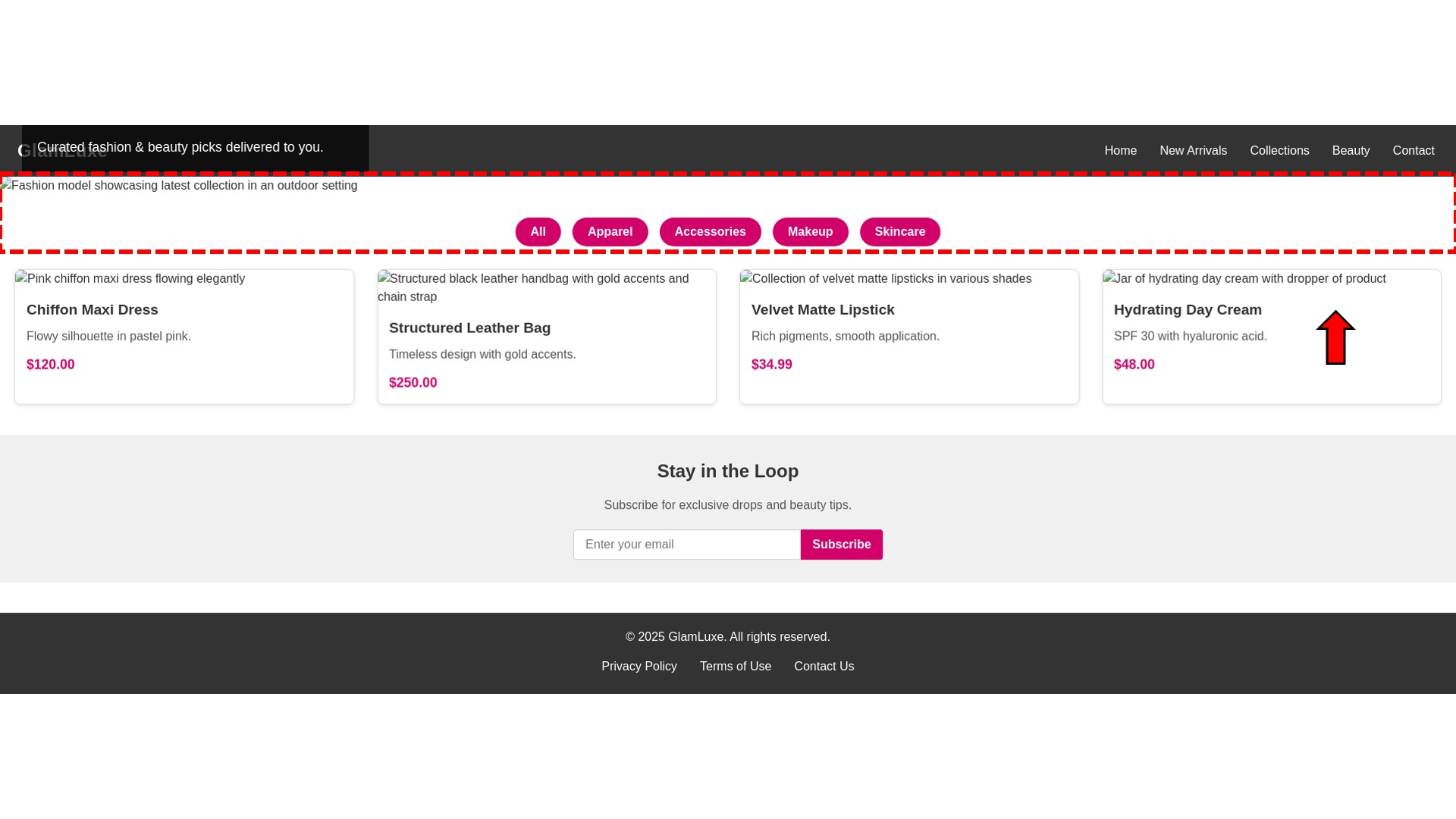}} \\
        GPT-5 & \bf Ours\\
        \frame{\includegraphics[width=0.47\linewidth]{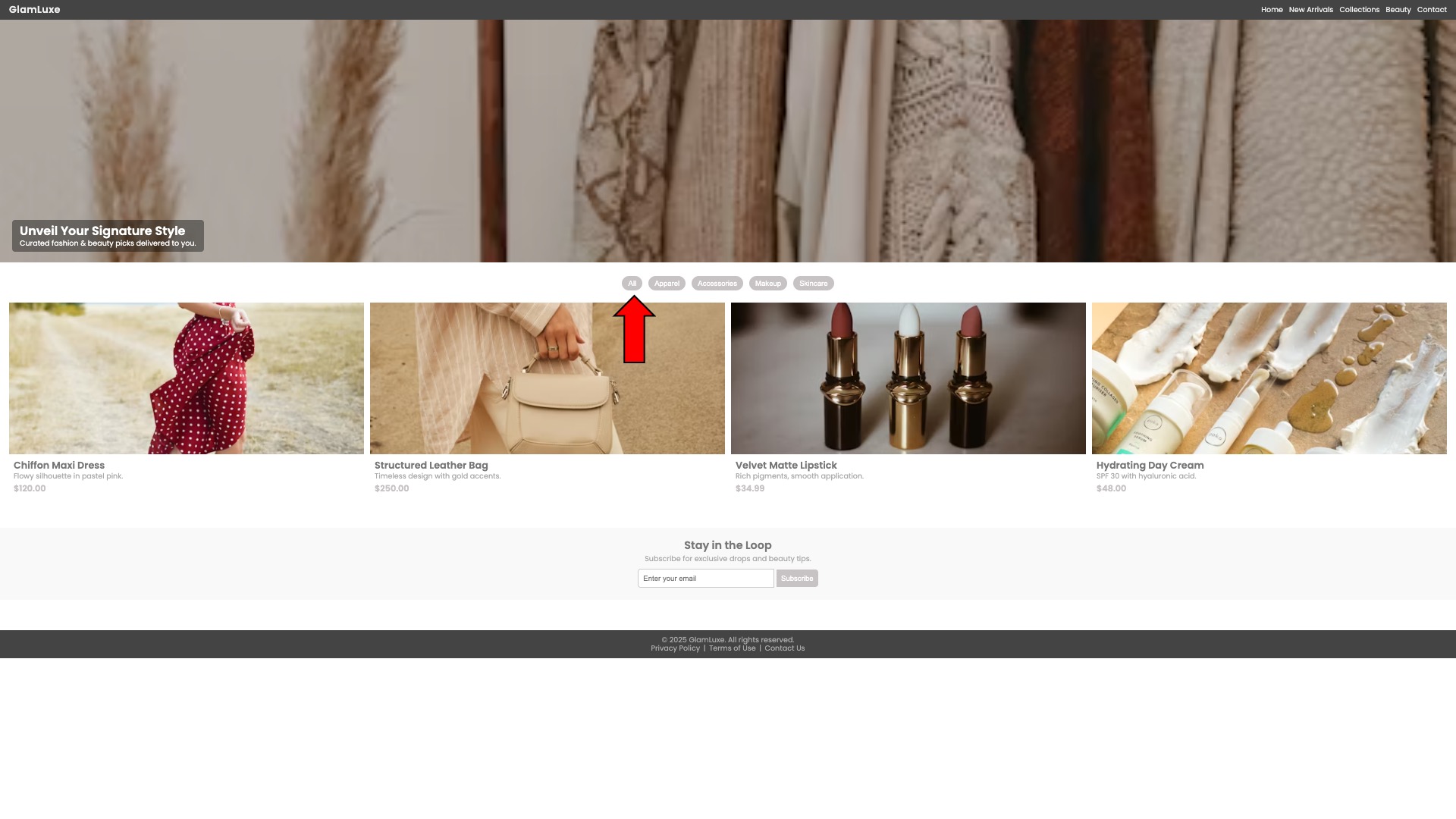}}  &      \frame{\includegraphics[width=0.47\linewidth]{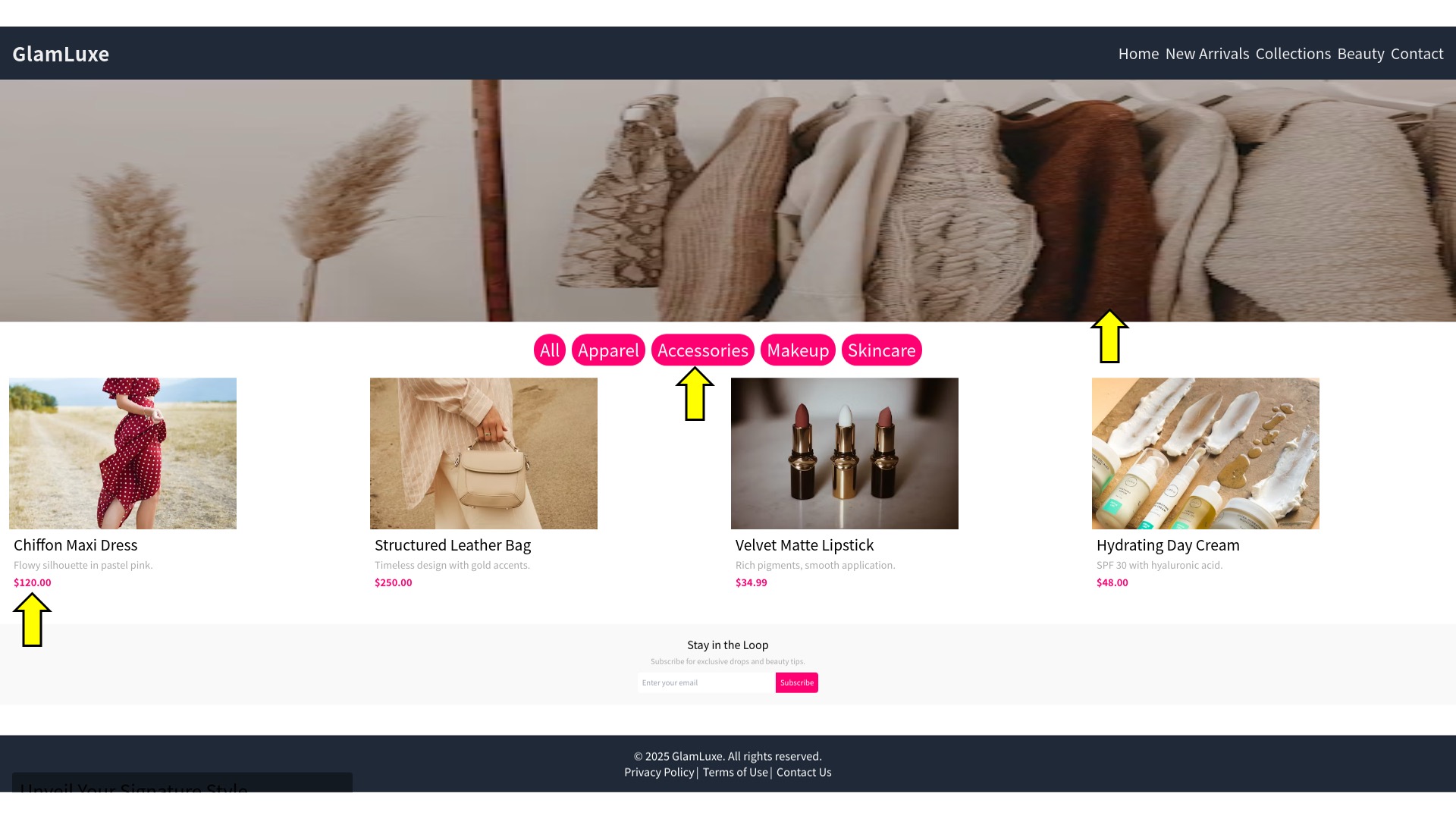}}\\
\end{tabular}        
    \caption{Yellow arrows highlight the fixed components, and red arrows show inaccessible parts. Ours shows it can keep the style while correcting inaccessible elements by changing their color contrast values. GPT-5 failed to put a correct contrast to buttons. Claude 3.5 failed to preserve the original web design by making the header and footer thicker. Also, it could not put the correct image URLs. Please zoom in for details.
    }
    \label{fig:appendix_q6}
\end{figure*}

\begin{figure*}[ht]
\centering
\begin{tabular}{ccc>{\columncolor{mygray}}c}
 Input & Claude3.5 \\ 
         \frame{\includegraphics[width=0.47\linewidth]{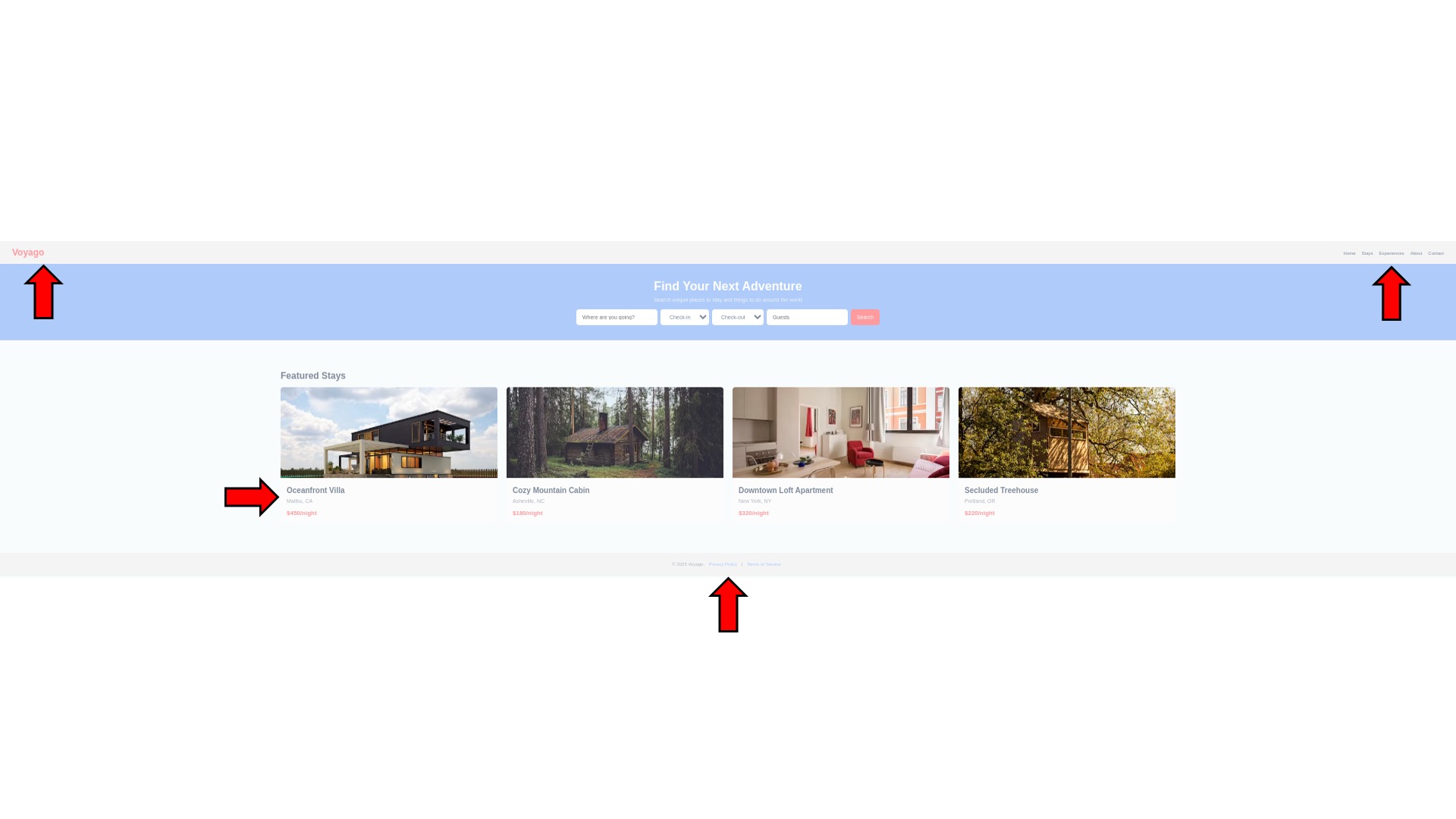}} &
        \frame{\includegraphics[width=0.47\linewidth]{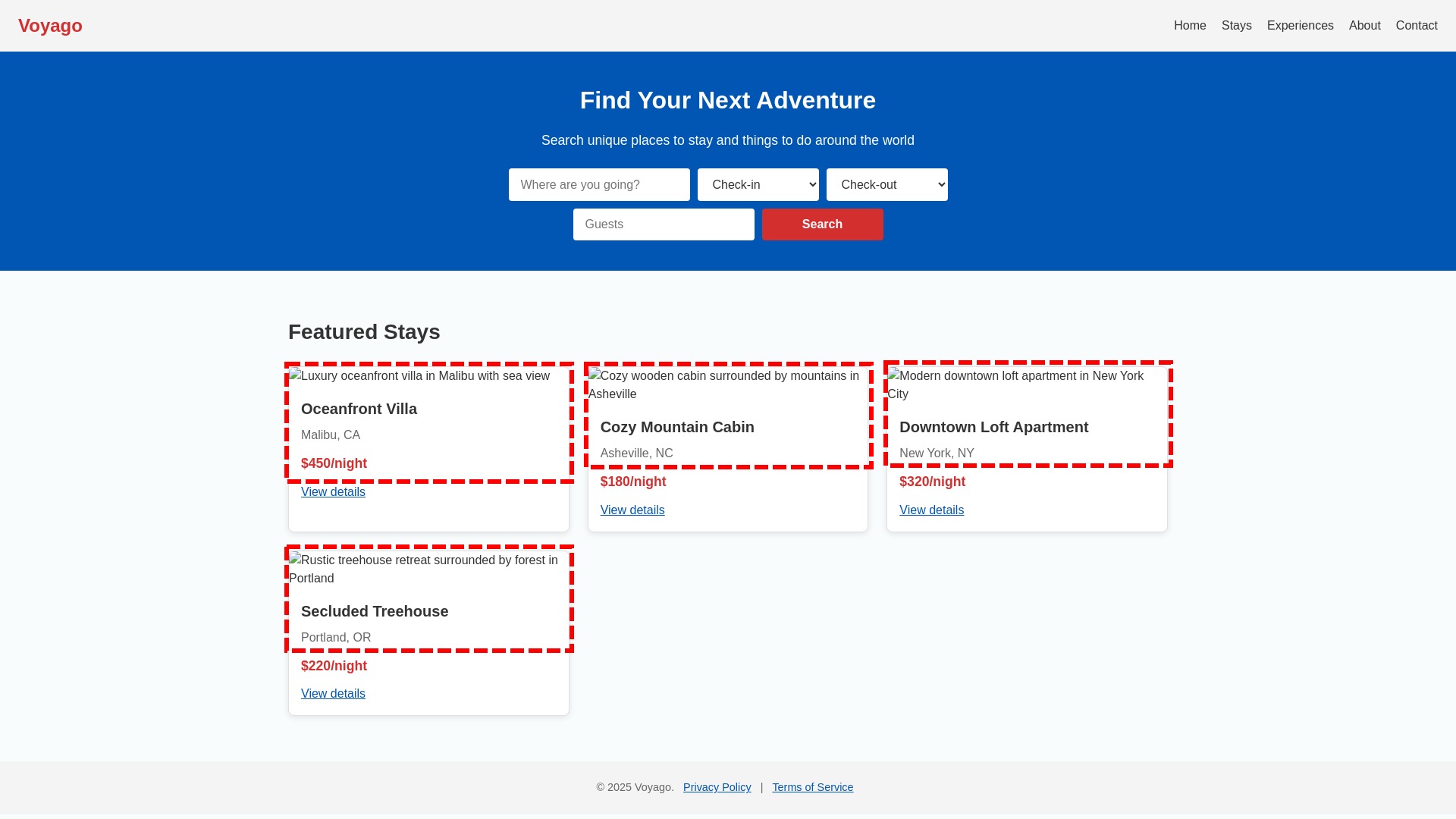}} \\
        GPT-5 & \bf Ours\\
        \frame{\includegraphics[width=0.47\linewidth]{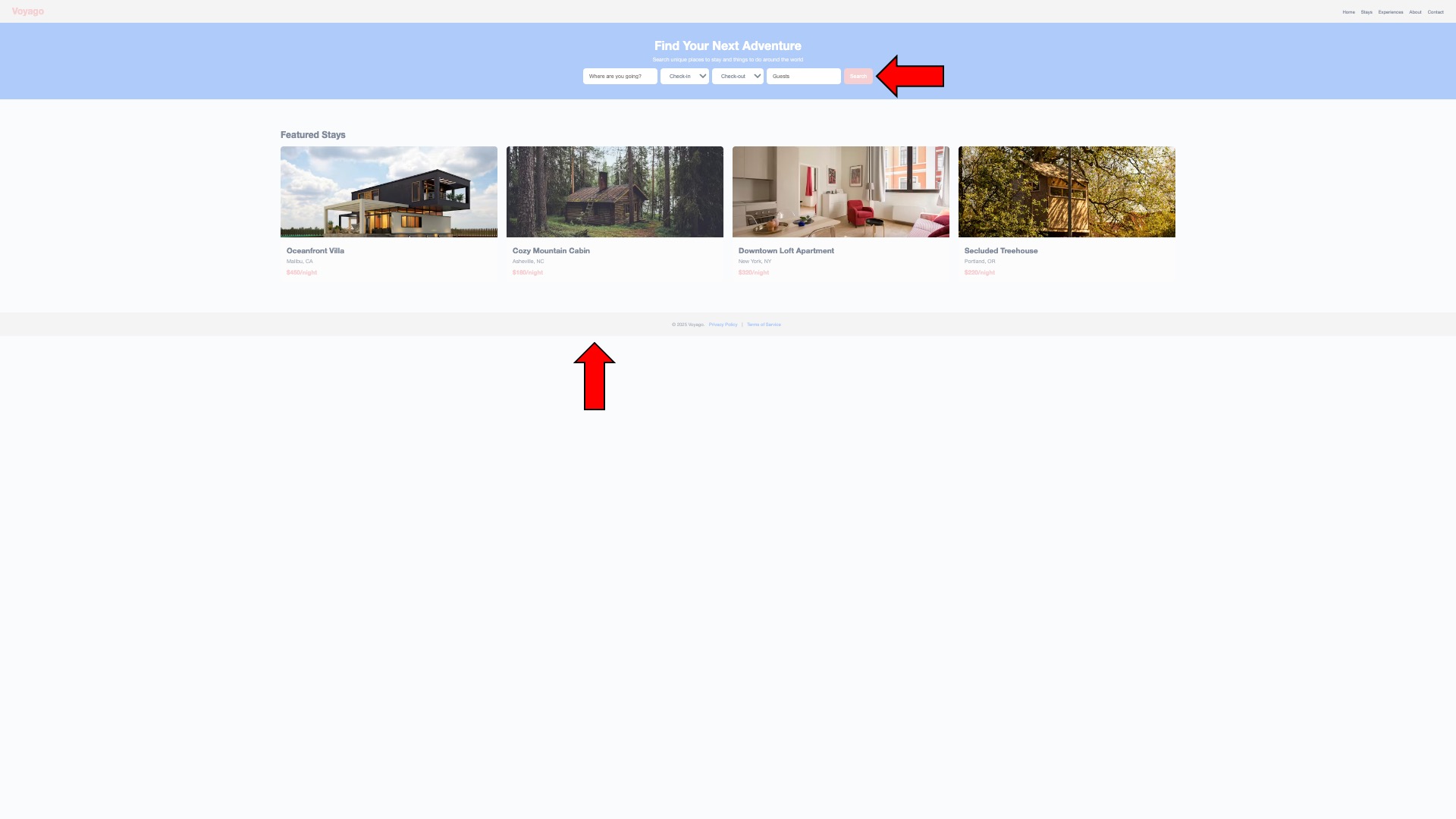}}  &      \frame{\includegraphics[width=0.47\linewidth]{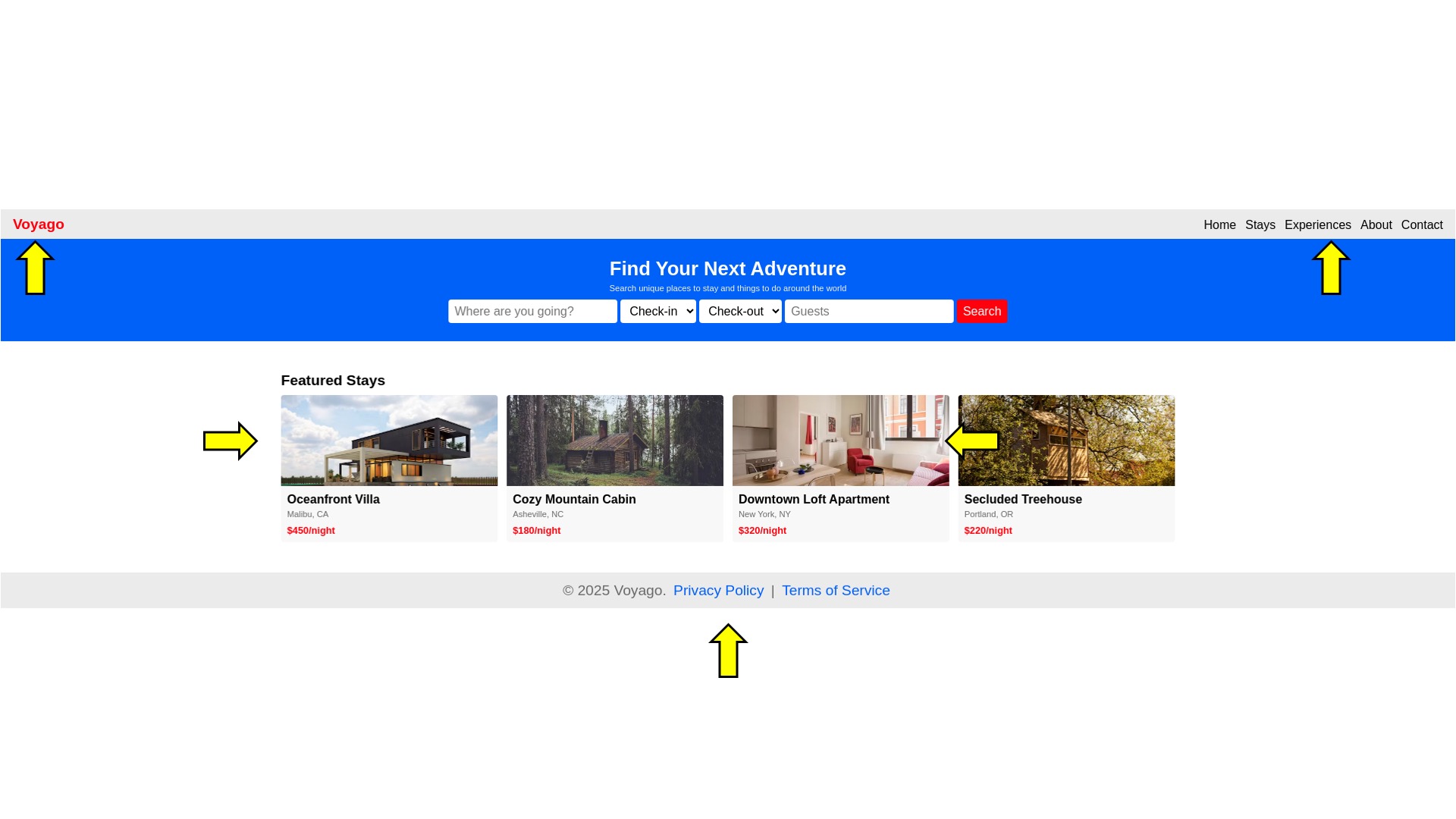}}\\
\end{tabular}        
    \caption{Yellow arrows highlight the fixed components, and red arrows show inaccessible parts. Ours shows it fixes text contrast in the header, image cards, and the footer by following the original web style. Claude 3.5  did not preserve the original style by making larger selection boxes and missing images. GPT-5 failed to put a correct contrast on the button and the footer. Please zoom in for details.
    }
    \label{fig:appendix_q7}
\end{figure*}

We show additional qualitative results using our violation-conditioned model, Claude 3.5 Sonnet~\cite{claude35sonnet}, and GPT-5~\cite{GPT5} in~\figref{fig:appendix_q1}--\figref{fig:appendix_q2} and~\figref{fig:appendix_q4}--\figref{fig:appendix_q7}.
\FloatBarrier

\FloatBarrier

\section{Ablation Study}
\label{sec:ablation_study}

\myparagraph{Violation conditioning: binary vs. report.}
We compare two conditioning strategies for guiding the VLM: (1) \textit{Binary}, which conditions the model on a binary indicator of whether violations exist (zero/non-zero), and (2) \textit{Report}, which conditions on the full structured violation report $\vc$ containing detailed violation types, messages, and locations. As shown in~\tabref{tab:ablation_module}, conditioning on the full violation report (Report) achieves fewer violations than binary conditioning (LLaVA~1.6: 0.366 vs.\ 0.509; Llama~3.2~Vision: 0.244 vs.\ 0.439), but results in substantially more aggressive structural modifications, as reflected by much higher tree edit distance (LLaVA~1.6: 11.141 vs.\ 3.053; Llama~3.2~Vision: 11.642 vs.\ 3.021). Instead, binary conditioning makes more conservative edits with lower tree edit distance, limiting unintended structural side-effects. Therefore, we use binary violation conditioning in our main experiments to balance accessibility improvement with structural fidelity.
\begin{table*}[ht]

    \centering
    
    \resizebox{\textwidth}{!}{
    \begin{tabular}{ccccccc}
        \specialrule{.15em}{.05em}{.05em}
        \textbf{Model} & Conditioning & Violation$\downarrow$ (\%)$\uparrow$  &  Caption-Img CLIP $\uparrow$ & SSIM $\uparrow$ &Tree Edit Dist.$\downarrow$\\
        \midrule
        \multirow{2}{*}{LLaVA 1.6} &  Binary &0.509    & 0.242 & 0.300 & 3.053 \\
        & Report &  0.366 & 0.236 & 0.910 & 11.141 \\ \hline
        \multirow{2}{*}{Llama 3.2 VL} & Binary &0.439    & 0.260 & 0.331 & 3.021 \\
        &  Report & 0.244 & 0.241 & 0.900 & 11.642 \\
       \specialrule{.15em}{.05em}{.05em}
    \end{tabular}
    }
    \caption{\textbf{Ablation study on violation conditioning strategies.} We compare conditioning on violation count (zero/non-zero) versus full violation report (detailed violation types and locations). Violation count conditioning achieves better design preservation (higher SSIM, lower tree edit distance) while maintaining low violation counts, demonstrating a favorable balance between accessibility improvement and structural fidelity.} 
    \label{tab:ablation_module}

\end{table*}

\myparagraph{Negative guidance ablation.} We analyze the impact of negative guidance sampling combined with the checker-in-the-loop mechanism across three VLMs. As shown in~\tabref{tab:ablation_negative_guidance}, negative guidance consistently reduces violations across all models. For LLaVA~1.6, violations decrease from 0.557 (Pass 1 without negative guidance) to 0.517 (Pass 1 with negative guidance), and further to 0.366 after the checker-in-the-loop refinement. Similar trends are observed for Llama~3.2~Vision (0.364 $\rightarrow$ 0.337 $\rightarrow$ 0.244) and Gemma~3 (0.370 $\rightarrow$ 0.352 $\rightarrow$ 0.234). Notably, the checker-in-the-loop mechanism provides additional refinement with minimal overhead, as only 10.7\%--18.9\% of files require Pass~2 (see~\tabref{tab:loop_efficiency}). The combination of negative guidance and iterative refinement achieves the best performance, reducing violations by 33.0\%--36.8\% compared to the initial Pass~1 without negative guidance, while maintaining high structural accuracy across all models.
\begin{table}[t]
    \centering
    \begin{tabular}{lccccc}
        \toprule
        \multirow{2}{*}{\textbf{Model}} & \multicolumn{2}{c}{\textbf{W/O Negative Guidance}} & \multicolumn{2}{c}{\textbf{W/ Negative Guidance}} \\
        \cmidrule(lr){2-3} \cmidrule(lr){4-5}
        & W/O Loop & W/ Loop & W/O Loop & W/ Loop \\
        \midrule
        LLaVA 1.6 & 0.557 & 0.381 & 0.517 & 0.366 \\
        Llama 3.2 Vision & 0.364 & 0.285 & 0.337 & 0.244 \\
        Gemma 3 & 0.370 & 0.259 & 0.352 & 0.234 \\
        \bottomrule
    \end{tabular}
    \caption{\textbf{Ablation study on negative guidance sampling with checker-in-the-loop.} We compare violation counts with and without negative guidance across multiple passes. W/O Loop shows results after the first generation, while W/ Loop shows the best result after up to $K=2$ passes. Negative guidance consistently reduces violations, particularly in the first pass, demonstrating its effectiveness in improving accessibility compliance.}
    \label{tab:ablation_negative_guidance}
\end{table}

\section{Computational Efficiency of Checker-in-the-Loop}
\label{sec:loop_efficiency}
\begin{table}[h]
    \centering
    \begin{tabular}{lc}
        \toprule
        \textbf{Model} & \textbf{Files Requiring Second Pass}  \\
        \midrule
        LLaVA 1.6 & 18.9\%  \\
        Llama 3.2 Vision & 11.4\%  \\
        Gemma 3 & 10.7\% \\
        \bottomrule
    \end{tabular}
    \caption{\textbf{Computational overhead of checker-in-the-loop.} We report the percentage of files requiring a second pass in our iterative refinement process (up to $K=2$ passes).}
    \label{tab:loop_efficiency}
\end{table}

To demonstrate that our checker-in-the-loop approach achieves substantial improvement with minimal computational overhead, we analyze the percentage of files requiring a second pass across different models. As shown in~\tabref{tab:loop_efficiency}, only 10.7\%--18.9\% of files require Pass 2 across all models, with Gemma~3 requiring the fewest at 10.7\%. This indicates that the vast majority of websites (over 81\%) converge to acceptable accessibility standards after just the first generation. Combined with the violation reduction shown in~\tabref{tab:ablation_negative_guidance}, these results demonstrate that our iterative refinement mechanism provides significant quality improvements while imposing minimal additional computational cost in practice.

\section{Implementation Details \& Prompt Ablation}
\label{supp:implementation_details}
\begin{table}[h]
    \centering

    \begin{tabular}{ccc}
    \specialrule{.15em}{.05em}{.05em}
       Model  & Trainable Module & Trainable \% \\ \hline
       Qwen 2 VL  & QKVO, Vision Proj, Vision Resampler  & 6.08 \\
       Qwen 2.5 VL & QKVO, Vision Proj, Vision Resampler  & 6.08 \\
       Qwen 3 VL & QKVO, Vision Proj, Vision Resampler  & 6.08 \\
       Llama 3.2 Vision & QV, Down, Up, Gate Proj & 22.06 \\
       LLaVA 1.5 & QKVO, Vision Proj & 13.90 \\
       LLaVA 1.6 & QKVO, FC1\&2, Layer Norm & 10.90 \\
       Gemma 3 & QV, Down, Up, Gate Proj & 0.27 \\
    \specialrule{.15em}{.05em}{.05em}
    \end{tabular}
    \caption{Trainable module and trainable parameter percentage of Violation-conditioned VLM.
    }
    \label{tab:trainable_param_stat}
\end{table}

\myparagraph{Training details.}
The trainable module and the trainable parameter percentage of each model are provided in~\tabref{tab:trainable_param_stat}.
Across all models, we train LoRA~\cite{hu2022lora} with a rank of $r=8$ and $\alpha=16$, resulting in a scaling factor of $\alpha / r = 2$. The dropout rate is set to 0.05. Due to memory constraints, we use gradient accumulation to enable training with a larger batch size. Specifically, we train Llama 3.2 Vision~\cite{dubey2024llama} with a batch size of 16,  LLaVA~1.6~\cite{liu2024llava16} with a batch size of 24, and Gemma 3 with a batch size of 13, all using negative guidance. The number of epochs for all the training runs is 100.
During training, the input text prompt is:
\begin{verbatim}[frame=single]
The following HTML was checked for WCAG 2.2 accessibility compliance.
Violation information is unknown. / For each violation listed below, 
locate the offending element and apply the fix directly in the HTML. 
<Parsed Violation Report>

<cond> The expected output HTML has unknown violations.
Modify the HTML below. Return only the corrected HTML 
without explanations.

Input HTML:
<input HTML x>
\end{verbatim}

\myparagraph{Inference details.}
During inference, we generate the output HTML code using the same prompt structure as training. To apply negative guidance sampling, we perform two forward passes at each token generation step: one conditioned on the violation report $\vc$ and one with the unconditional prompt $\vc_{\emptyset}$. The logits from both passes are combined using the guidance scale $\gamma$ to compute the final token probabilities.

For checker-in-the-loop inference, we iteratively refine the output for a maximum of $K=2$ passes. At each iteration, we: (1) generate corrected HTML $\vy^{(i)}$ from the input conditioned on violations $\vc^{(i-1)}$, (2) run the accessibility checker on $\vy^{(i)}$ to obtain residual violations $\vc^{(i)}$, and (3) if violations remain and $i < K$, use $\vy^{(i)}$ and $\vc^{(i)}$ as input for the next iteration. We track the HTML with the lowest violation count across all iterations as the final output. In practice, most samples converge after the first or second pass.

\myparagraph{API calls.} 
We use a public API from OpenAI for ChatGPT4o with an image to run an inference using the following code as VLM. For LLM, we did not send the image data as the official document introduced. For all publicly available models, including Gemini 1.5~\cite{team2024gemini} and Claude 3.5~\cite{claude35sonnet}, we use the same prompt but different API calls using their API functions. 

\begin{lstlisting}[language=Python]
import base64
from openai import OpenAI

# Function to encode the image
def encode_image(image_path):
    with open(image_path, "rb") as image_file:
        return base64.b64encode(image_file.read()).decode("utf-8")

        
client = OpenAI()
# HTML = HTML Path.
# image_path = A rendering from the HTML.

base64_image = encode_image(image_path)
with open(html, "r", encoding="utf-8") as file:
    html_content = file.read()
    
response = client.chat.completions.create(
    model="gpt-4o",
    messages=[
        {
            "role": "user",
            "content": [
                {
                    "type": "text",
                    "text": f"Modify the following HTML code to comply with
                    accessibility guidelines, WCAG 2.2, and only output the
                    HTML code without explanations. {html_content}",
                },
                {
                    "type": "image_url",
                    "image_url": {"url": f"data:image/jpeg;base64,{base64_image}"},
                },
            ],
        }
    ],
    )

\end{lstlisting}

\myparagraph{Refined prompt experiments.}
We further examine whether a higher-quality prompt can improve the performance of baseline models. Following the OpenAI Prompt Engineering Guide, we redesigned the prompt to include explicit context, detailed instructions, and concrete accessibility requirements. The full prompt is:

\begin{verbatim}
Context:
You are a professional software engineer specializing in improving web
accessibility and WCAG 2.2 compliance. You are tasked with auditing and
fixing HTML source code to ensure it meets accessibility standards.

Instructions:
- Review and apply your comprehensive knowledge of WCAG 2.1.
- Fix accessibility violations in the provided HTML code.
- Maintain the original visual and interactive design as closely as possible.
- Prioritize interactive elements that require user input or navigation.
- Ensure semantic HTML is used correctly (e.g., headings hierarchy,
  <main>, <nav>, <section>, etc.).
- Only add ARIA roles or attributes when necessary.
- All <img> tags must have descriptive alt attributes.
- Ensure color contrast meets WCAG AA.
- All interactive elements must be keyboard accessible,
  focusable, and correctly labeled.
- Provide <label> elements for form inputs.
- Ensure <html> has a correct lang attribute and include a meaningful <title>.

{HTML Code goes here}
\end{verbatim}
Our previous prompt simply stated:
\begin{verbatim}
Modify the HTML code to comply with WCAG 2.2 Accessibility Guidelines.
{HTML Code goes here}
\end{verbatim}

We evaluate both prompts using the same test set and report the number of accessibility violations after correction. As shown in \tabref{tab:prompt_refinement}, the refined prompt improves the performance of baseline models. For example, ChatGPT-4o reduces violations from 2.929 to 1.793. However, our method remains substantially more effective, achieving only 0.439 violations.

\begin{table}[h]
\centering
\begin{tabular}{lc}
\toprule
Model & Number of Violations \\
\midrule
ChatGPT-4o (Refined prompt) & 1.793 \\
ChatGPT-4o (Basic prompt) & 2.929 \\
Gemini-2.0-flash (Refined prompt) & 2.134 \\
Gemini-1.5-flash (Basic prompt) & 6.213 \\
\midrule
Ours & \textbf{0.439} \\
\bottomrule
\end{tabular}
\caption{Remaining accessibility violations after correction under different prompt designs.}
\label{tab:prompt_refinement}
\end{table}

\myparagraph{Impact of visual input.} 
To verify that the visual input is not ignored by the model, we conduct an experiment in which the rendered image is replaced with a solid black image while leaving the HTML code unchanged. Under this setting, LLaVA~1.6 shows a clear drop in performance, with violations rising from 0.527 to 0.832, a 56.9\% increase. This indicates that the model does use visual cues to detect and fix accessibility errors.

\section{Test Statistics of User Study}
\label{sec:test_stat}

Let $n$ be the total number of valid votes (skipped answers are excluded), and let $n_i$ be the number of votes for method $i\in\{A,B,C\}$. Under $H_0^{\mathrm{global}}: p_A=p_B=p_C$, the expected count for each method is
\[
E_i \;=\; n\,p_0,\qquad p_0=\tfrac{1}{3}.
\]
The Pearson chi-squared statistic is
\[
\chi^2 \;=\; \sum_{i\in\{A,B,C\}} \frac{(n_i - E_i)^2}{E_i}
\;=\; \sum_{i\in\{A,B,C\}} \frac{(n_i - n/3)^2}{\,n/3\,},
\]
which follows a $\chi^2$ distribution with $2$ degrees of freedom under $H_0^{\mathrm{global}}$. We report the associated $p$-value and Cramér's $V=\sqrt{\chi^2/(n\,(k-1))}$ with $k=3$ as an effect size.

\section{Training and Inference Time}
\label{sec:train_inf_time}
We report the training and inference time of the supervised fine-tuned VLMs in~\tabref{tab:training_inference_time}. Specifically, we provide the training time for one epoch and the average inference time for a single HTML input. As expected, the training time remains unchanged with or without our violation-conditioning technique. Additionally, the inference time matches the corresponding pre-trained models, as our approach does not introduce new model parameters.

\begin{table}[h]
    \centering

    \begin{tabular}{ccc}
    \specialrule{.15em}{.05em}{.05em}
       Model  & Training Time (sec.) & Inference Time (sec.) \\ \hline
       Qwen 2 VL  & 168.6  & 3.08 \\
       Qwen 2.5 VL & 187.8  & 3.20 \\
       Llama 3.2 Vision & 368.4 & 7.32 \\
       + Neg. Guidance & - & 9.40 \\
       LLaVA 1.5 & 144.0 & 10.50 \\
       LLaVA 1.6 & 202.8 & 5.64 \\
       + Neg. Guidance & - & 9.30 \\
    \specialrule{.15em}{.05em}{.05em}
    \end{tabular}
    \caption{Training time of VLMs fine-tuned on \textit{WebAccessVL} for one epoch and average inference time for one HTML code.
    }
    \label{tab:training_inference_time}
\end{table}

\section{Website Correction Process for \textit{WebAccessVL}}
\label{sec:website_correction}
{
We introduce how we manually correct a website to comply with WCAG guidelines, as we introduced in~\secref{sec:dataset}. First, we load the HTML code and visually check the content to preserve its design style.
We run the IBM checker~\cite{checker} on the HTML code to check violations using the following command:}
{\footnotesize
\begin{Verbatim}[frame=single]
achecker --outputFolder output myHTML.html
\end{Verbatim}
}
{
It will output violations, if any, and their XPath to locate in the HTML code, along with their snippet. It also includes an IBM website link for a brief solution and its message to explain the violation, like the following:
}
{\footnotesize
\begin{Verbatim}[frame=single]
Level: violation
XPath: /html[1]/body[1]/div[1]/div[2]/table[1]/tbody[1]/tr[2]/td[3]
Snippet: <td aria-colindex="2" role="cell">
Help: https://able.ibm.com/rules/archives/...
- Message: An explicit ARIA 'role' is not valid for <tr> element 
  within an ARIA role 'table' per the ARIA in HTML specification
\end{Verbatim}
}
{
Next, we go to the XPath that is flagged as the violation, and we fix it to minimize the changes in the original design. We then visually inspect the updated website to ensure there are no significant structural or color changes. While reducing the number of violations is important, preserving the original design remains a key priority.
We repeat this process until all violations are resolved or until the remaining violations require significant modifications to the existing design.
On average, fixing a single website takes 7 to 10 minutes. Constructing \textit{WebAccessVL}, which contains 2500 manually corrected websites, therefore requiring approximately 417 hours in total.
}

\section{Broader Impact}
\label{sec:broader_impact}
This work contributes to improving web accessibility by using vision-language models to automatically reduce WCAG 2.2 violations in HTML. It has the potential to benefit users with disabilities and support developers by lowering the technical barrier to creating accessible websites. By relying on open-weight models, our approach also offers a privacy-conscious alternative to commercial services. However, automated fixes may optimize for rule-based compliance without ensuring real usability, and over-reliance on such systems could reduce the incentive for inclusive design practices. We view this method as a tool to support, not replace, human developers and encourage continued collaboration with accessibility experts and users.

\end{document}